\documentclass[prb,superscriptaddress,letterpaper,floatfix,preprintnumbers,showpacs]{revtex4-1}
\usepackage{amsfonts}
\usepackage{amsmath}
\usepackage{amssymb}
\usepackage{bm}
\usepackage{color}
\usepackage{dcolumn}	% Align table columns on decimal point
\usepackage{gensymb}
\usepackage{graphicx}	% Include figure files
\usepackage[hyperfootnotes,hyperindex,hyperfigures]{hyperref}
\usepackage{mathcomp}
\usepackage{textcomp}

\providecommand{\by}{$\times$}
 
\definecolor{adobe}{rgb}{.8,.6,.5}
\definecolor{mygreen}{rgb}{0.0,0.6,0.0}
\definecolor{blue2}{rgb}{0.0,0.0,0.8}
\definecolor{brown}{rgb}{0.6,0.3,0.0}
\definecolor{forest}{rgb}{0.0,0.4,0.0}
\definecolor{grass}{rgb}{0.0,0.55,0.25}
\definecolor{grass2}{rgb}{0.0,0.6,0.25}
\definecolor{gray}{rgb}{0.4,0.4,0.4}
\definecolor{grayish}{rgb}{0.2,0.2,0.4}
\definecolor{khaki}{rgb}{0.9,0.9,0.7}
\definecolor{lightteal}{rgb}{0.0,0.6,0.6}
\definecolor{lightyellow}{rgb}{1.0,1.0,0.5}
\definecolor{maroon}{rgb}{0.7,0.1,0.2}
\definecolor{navy}{rgb}{0.0,0.1,0.7}
\definecolor{olive}{rgb}{0.4,0.4,0.0}
\definecolor{orange}{rgb}{0.9,0.45,0.0}
\definecolor{peach}{rgb}{1.0,.8,.7}
\definecolor{purple}{rgb}{0.4,0,0.55}
\definecolor{teal}{rgb}{0.0,0.5,0.4}
\definecolor{turq}{rgb}{0.3,0.6,0.9}
\definecolor{violet}{rgb}{0.75,0,0.75}
\hypersetup{colorlinks=true, linkcolor=navy, citecolor=blue2}% linkcolor=FIGURE numbers citecolor=citation colors

\begin{document}

%\preprint{UPDATED: {\color{maroon} November 13, 2011}}

%: TITLE OF PAPER +++++++++++++++++++++++++++++++++++++++++++++++++++++++++++

\title{{\color{navy} Improved Semiconductor Lattice Parameters and Band Gaps from a Middle-Range Screened Hybrid Exchange Functional}}

%: AUTHOR LIST ++++++++++++++++++++++++++++++++++++++++++++++++++++++++++++++++
\author{Melissa J. Lucero}
\affiliation{Department of Chemistry, Rice University, Houston, TX 77005-1892, USA}

\author{Thomas M. Henderson}
\affiliation{Department of Chemistry, Rice University, Houston, TX 77005-1892, USA}
\affiliation{Department of Physics \& Astronomy, Rice University, Houston, Texas 77005-1827, USA}

\author{Gustavo E. Scuseria*}
%\email{Electronic mail: guscus@rice.edu}
%\homepage{http://python.rice.edu/~guscus/}
\affiliation{Department of Chemistry, Rice University, Houston, TX 77005-1892, USA}
\affiliation{Department of Physics \& Astronomy, Rice University, Houston, Texas 77005-1827, USA}
\affiliation{Chemistry Department, Faculty of Science, King Abdulaziz University, Jeddah 21589, Saudi Arabia}

%: ABSTRACT ++++++++++++++++++++++++++++++++++++++++++++++++++++++++++++++++  

\begin{abstract}

We show that the middle-range exchange-correlation hybrid of Henderson, Izmaylov, Scuseria and Savin (HISS) performs extremely well for elemental and binary semiconductors with narrow or visible spectrum band gaps, as well as some wider gap or more ionic systems used commercially. 
The lattice parameters are superior to those predicted by the screened hybrid functional of Heyd, Scuseria and Ernzerhof (HSE), and provide a significant improvement over geometries predicted by semilocal functionals. 
HISS also yields band gaps superior to those produced by functionals developed specifically for the solid state.
Timings indicate that HISS is more computationally efficient than HSE, implying that the high quality lattice constants coupled with improved optical band gap predictions render HISS a useful adjunct to HSE in the modeling of geometry-sensitive semiconductors.
\end{abstract}

 %	Optical properties of bulk materialsand thin films

\maketitle

% ||||||||||||||||||||||||||||||||||||||||||||||||||||||||||||||||||||||||||||||||||||||||||||||||||||||||||||||||||||||||||||||||||||||||||||||||||||||||||||||||||||
%					 |||||||||||||||||||||||||||||||||||||||||||  MANUSCRIPT BEGINS   |||||||||||||||||||||||||||||||||||||||||||||||||||||||||||
% ||||||||||||||||||||||||||||||||||||||||||||||||||||||||||||||||||||||||||||||||||||||||||||||||||||||||||||||||||||||||||||||||||||||||||||||||||||||||||||||||||||

%|||||||||||||||||||||||||||||||||||||||||||
\section{Introduction} \label{sec:Intro}%====================================================
%|||||||||||||||||||||||||||||||||||||||||||
%
Reliable theoretical treatment of semiconductors,  particularly those in the visible range (20 meV-4 eV) remains non-trivial.
While semilocal DFT methods are the most tractable approach in terms of computational expense, the band gaps obtained are often not that impressive.\cite{Staroverov2005,Dreuw2006,Zhao2006,Peach2008}
Moreover, many semiconductors, such as the perovskites,\cite{Perovskites1,Iles2010} chalcopyrites\cite{Chalcopyrites2,Chalcopyrites1,Chalcopyrites3} or Delafossites\cite{Delafossites2,Delafossites1} have solid state properties that are very sensitive to geometry.
These shortcomings of solid state DFT arise, in part, from the various approximations of the exchange-correlation energy, $E_{xc}$.\cite{JacobsLadder}

In the simplest case, the Local Spin Density Approximation (LSDA)\cite{LSDA} assumes that the exchange-correlation energy density at the point $\bm{r}$ depends only on the spin densities at  $\bm{r}$, rendering the exchange interaction \emph{local}.\cite{Perdew2001} 
For solids, the LSDA underestimates lattice constants,\cite{Khein1995,Kurth1999} and severely underestimates band gaps.\cite{Perdew1986}
Generalized gradient approximations (GGAs) such as the functional of Perdew, Burke, and Ernzerhof (PBE),\cite{PBE,Perdew1997} improve upon the LSDA by incorporating the local density gradient, while the meta-GGAs such as the functional of Tao, Perdew, Staroverov, and Scuseria (TPSS)\cite{TPSS} also add dependence on the kinetic energy density and/or the Laplacian of the density.  
While GGAs and meta-GGAs are \textit{semilocal} in character, they can partially account for inhomgeneities in the density, thus showing significant improvement over the LSDA in atoms and molecules, but for extended systems, these functionals can often provide predictions of lower quality than the LSDA.\cite{Ma1968,Perdew1986b,Perdew1986c}

Hybrid functionals including a percentage of exact \textit{non-local} Hartree-Fock-type exchange\cite{Becke1993a,Becke1993b} have proven very successful for molecules, as is evident from the popularity of the B3LYP\cite{B3LYP} functional.  
These hybrids describe the exchange-correlation energy as
	\begin{equation}\label{eq:GKS}
		E_{xc} = E^{DFA}_{xc} + c \left(E^{HF}_x - E^{DFA}_x\right)
	\end{equation}
where $E^{DFA}_{xc}$ is the exchange-correlation energy of a (semi)local density functional approximation (DFA), $E_x^{HF}$ and $E_x^{DFA}$ are, respectively, the exact exchange energy and the exchange energy from a DFA and, the constant $c$ typically ranges from $1/10$ to $1/2$.  
Unfortunately, even increasingly sophisticated hybrids such as PBEh,\cite{PBE0,PBE1PBE} TPSSh,\cite{TPSSh} or the highly empirical M06\cite{M062X} series of functionals have been unable to supplant the LSDA for applications in extended systems, despite its numerous and well-known insufficiencies.  

In part, this failure is a matter of computational affordability, as the lattice sums required to evaluate the non-local exchange energy are slowly convergent, particularly as the band gap of the system decreases.
A second problem arises from the fact that hybrid functionals are almost always constructed in the generalized Kohn-Sham (GKS) framework,\cite{GKS} where the the nonlocal exchange energy leads to a non-local exchange ``potential'' similar to that of Hartree-Fock (HF).  
Accordingly, hybrid functionals formulated in the GKS sense inherit many of the deficiencies that Hartree-Fock exhibits for extended systems. 
In fact, for some solid-state properties, hybrid functionals yield errors similar in magnitude to those of semilocal functionals, but at much greater computational cost.\cite{RangeSep,Paier2006}

In order to efficiently recover the benefits of non-local exchange for both solids and molecules, a more flexible scheme to mix exact and semilocal exchange is required.  
Probably the most promising route is the use of range-separation,\cite{SavinBook,Savin1995,Savin1997,RangeSep} where the amount of non-local exchange included in the functional depends on the distance between electrons.  
A typical range-separation scheme might split the Coulomb operator $1/r$ into short range (SR) and long range (LR) pieces, as
	\begin{equation}\label{eq:RangeSep}
		\frac{1}{r}=\underbrace{\frac{1 - \textrm{erf}(\omega r)}{r}}_{SR} + \underbrace{\frac{\textrm{erf}(\omega r)}{r}}_{LR},
	\end{equation}
where $\omega$ is a parameter defining the length scale of separation and $\textrm{erf}$ is the error function, chosen for simplicity of integration with gaussian orbitals. 
The exchange-correlation energy then becomes
	\begin{equation}
		\begin{split}
		E_{xc} = E^{DFA}_{xc} &+ c_{SR} \left(E^{SR-HF}_x - E^{SR-DFA}_x\right)
		\\
                  & + c_{LR} \left(E^{LR-HF}_x - E^{LR-DFA}_x\right)
		\end{split}
	\end{equation}
so that the amount of non-local exchange used for electrons which are far apart can differ from the amount used for electrons which are close together.  
In molecules, we generally set $c_{LR} = 1$ in what is known as a long-range-corrected hybrid.  
By choosing $c_{LR} = 1$, the correct long-range portion of the exchange-correlation potential is obtained, resulting in substantial improvements over non-local hybrids for properties which sample density tails, including non-linear optical properties,\cite{LCwPBE} long-chain polarizabilities,\cite{Hirao2001} and charge transfer or Rydberg excitations.\cite{Sekino2005}  
Solids, however, have no density tails, and the long-range portion of exact exchange becomes problematic if left unscreened, not only because it is expensive to compute, but because it yields unphysical results.  
We, therefore, set $c_{LR} = 0$ in what are known as screened hybrids.

The screened hybrid functional of Heyd, Scuseria, and Enzerhof\cite{HSE03,HSE06,HSEh} uses 25\% HF exchange in the short range, with $\omega = 0.11a_{0}^{-1}$, where $a_0$ is the Bohr radius.  
This value of $\omega$ corresponds to a range of $\sim$ 5 \AA, so that non-local exchange extends over two to three nearest neighbors.  
Note that for practical purposes this range is quite long.
The HSE screened hybrid works reasonably well for metals, but significantly improves upon the predictions of regular hybrids for semiconductors. 
For molecular systems, it yields results of the same quality as unscreened  hybrids. 
Nevertheless, despite some ability to describe long range properties,\cite{HSEh} HSE does not adequately describe quantities sensitive to the long-range portion of the exchange-correlation potential.\cite{Janesko2009,Barone2011}

A functional successfully coupling the accuracy for solids of a screened hybrid with the accuracy for molecules of a long-range-corrected hybrid would be of considerable utility in modeling extended systems, as surface phenomena such as molecular adsorption, catalysis, and surface reconstruction could then be reliably addressed.  
This was the motivation behind the development of the HISS functional of Henderson, Izmaylov, Scuseria, and Savin\cite{HISS1,HISS2} which has three ranges, rather than just two.  
Like HSE, HISS uses PBE as the underlying GGA, but HISS writes
	\begin{equation}\label{eq:HISSxc}
		E^{HISS}_{xc} = E^{PBE}_{xc} + c_{MR} \left(E^{MR-HF}_{x} - E^{MR-PBE}_{x}\right),
	\end{equation}
where the last two terms in parentheses are the middle-range (MR) exact exchange and middle-range PBE exchange energies, given by
	\begin{equation}
	\frac{1}{r}=
	\underbrace{\frac{\textrm{erfc}(\omega_{SR} r)}{r}}_{SR} + 
	\underbrace{\frac{\textrm{erf}(\omega_{LR} r)}{r}}_{LR} + 
	\underbrace{\frac{\textrm{erf}(\omega_{SR} r)-\textrm{erf}(\omega_{LR} r)}{r}}_{MR}.
\end{equation}
	
The three parameters were determined by fitting to small test sets of atomization energies, barrier heights, and band gaps, resulting in $\omega_{SR}=0.84a^{-1}_{0}$, $\omega_{LR}=0.20a^{-1}_{0}$, and $c_{MR}=0.60$.  
The fraction of exact exchange included in HISS rises from zero at $r=0$ to roughly 0.35 near $r = 1.5 a_0$, then decays to zero again for large distances; HISS has more exact exchange than does HSE over one or two chemical bonds, but the fraction of exact exchange decays even more rapidly than it does in HSE, producing a speed-up in solid-state calculations.

As expected, enhanced mid-range non-local exchange was demonstrated to reduce calculated errors for thermochemistry, low-lying Rydberg states, atomization energies, barrier heights and polarizabilities of H$_{2}$ chains.\cite{HISS2}  
For the semiconductor band gaps tested, HISS did not performs as well as HSE, overestimating them by 0.40 eV.\cite{HISS2}  
However, given the excellent reproduction of molecular properties and the fact that only a small set of solids was examined, we chose to more thoroughly assess the performance of HISS in semiconductors.

%|||||||||||||||||||||||||||||||||||||||||||
\section{Computational Methods} \label{sec:Methods}% ---------------------------------------------------------
%|||||||||||||||||||||||||||||||||||||||||||

Density functional calculations were performed using the periodic boundary-condition code\cite{Kudin2000,Kudin1998b,Kudin1998a} within the {\sc{gaussian}} suite of programs.\cite{G09} 
Gaussian basis sets optimized for solids were discussed and provided in the Supplementary Material\cite{HSEepaps} of the earlier study by Heyd and co-workers.\cite{Heyd2005} 

Band gaps and related properties reported here are for fully-relaxed, equilibrium structures (lattice parameters and geometries) obtained using the LSDA (with SVWN5\cite{SVWN5}), PBE, TPSS, PBEsol,\cite{PBEsol} the non-local screened hybrid functional, HSE, and the middle-range screened hybrid functional HISS.
The equilibrium structures are optimized in redundant internal coordinates\cite{Kudin2001} using the {\sc{gaussian09}} defaults.
The SCF convergence criterion was set to $10^{-7}$ on the density matrix ({\sc{scf=tight}}) with Brillouin space integration performed using 12 points in each dimension, for a total of 868 points.
All extended systems studied are for 3-dimensional ``bulk'' crystalline materials.

%|||||||||||||||||||||||||||||||||||||||||||
\section{Benchmarks} \label{sec:Results}% ---------------------------------------------------------
%|||||||||||||||||||||||||||||||||||||||||||
%
%|||||||||||||||||||||||||||||||||||||||||||
\subsection{The SC40 Test Set} \label{subsec:sc40} % ====================================================
%|||||||||||||||||||||||||||||||||||||||||||
%

The solids used in this study were previously published as the SC40 test set.\cite{Heyd2005} 
The 40 compounds are here separated into three groups as presented in Table~\ref{tab:sc40}.
Group 1 contains twenty commercially-relevant bulk semiconductors absorbing in the optical (0.20 meV to 4 eV) range.
This set includes the most commonly used or benchmarked elemental and binary compounds adopting the diamond (di), zincblende (zb), or wurtzite (wu) lattices, and each compound is referenced to a complete set of  low (LT) and room temperature (RT) band gap data. 
Group 2 consists of a smaller set of wide-gap semiconductors and ionic solids known to be challenging computationally or for various intriguing experimental properties. 
Group 3 encompasses a collection of binary compounds useful for simultaneously benchmarking lattice parameters and providing an opportunity for band gap prediction as these materials have gaps that are either unknown or not particularly well-established.
Each group is further subdivided according to the lattice structure, with the diamond, zincblende and wurtzite structures preceding any rock salt (rs) structures within each group.

To facilitate evaluation of the relative performance of each functional, the experimental band gaps have been updated to reflect half a decade of experimental progress and further augmented to include, and discriminate between, room temperature (RT) and low temperature (LT) values.
It should be noted that the low temperature data set is not exhaustive and the experimental measurements range from extrapolated 0 K values up through 27 K, providing a rather inconsistent basis by which to judge the quality of the predicted optical gaps, particularly if the temperature dependence of a given material is uncertain. 
Alternatively, the more complete room temperature (RT) data set, which has gaps typically ca. 0.2 eV higher in energy than the measured LT results, can provide a useful baseline for estimations of the low temperature gaps where unavailable or extrapolated.
The majority of the experimental reference data (gaps and lattice constants) are taken from the 3rd Edition of the Springer Semiconductor Handbook\cite{Madelung2004} and any references included therein. 
Alternate tabulations are cited where applicable.

%
%: TABLE Lattice Parameters and Band Gaps  TTTTTTTTTTTTTTTTTTTTTTTTTTTTTTTTTTTTTTTTTTTTTTTTT
\begin{table*}[!htb]\caption{\label{tab:sc40} Equilibrium lattice parameters (\AA) and band gaps (eV) at the relaxed geometries for the SC40 solid test set.\cite{HSEepaps,Heyd2005} Initial lattice constants are room temperature experimental values except for the six presented in Table~\ref{tab:sols}. No ZPAE corrections have been made. The reference set for band gaps is either low temperature (LT) or room temperature (RT). }
\begin{ruledtabular}
\begin{tabular}{l cc ccccc rcc c ccccc}
& \multicolumn{2}{c}{Expt. Lattice\footnotemark[1]} & \multicolumn{5}{c}{Calculated Lattice Parameters} &&\multicolumn{3}{c}{Expt. Gap\footnotemark[2] }&\multicolumn{5}{c}{Calculated Band Gaps} \\ 
	\cline{2-3} \cline{4-8} \cline{10-12} \cline{13-17}\noalign{\smallskip} 
Solid & Type & $a_{0}$ or $a, c$& HISS & HSE & LSDA &TPSS & PBEsol &&  LT & 300 K & Type & HISS & HSE & LSDA & TPSS & PBEsol\\
\hline
\multicolumn{10}{l}{\textbf{Group I: Semiconductors}} \\
Ge	&	di	&	5.658	&	5.660	&	5.706	&	5.650	&	5.743	&	5.693	
&&	0.74	&	0.66	&	I	&	1.08	&	0.54	&	0.00	&	0.00	&	0.00	\\
Si	&	di	&	5.430	&	5.430	&	5.451	&	5.410	&	5.465	&	5.442	
&&	1.17	&	1.12	&	I	&	1.45	&	1.22	&	0.51	&	0.74	&	0.53	\\
$\beta$-SiC	&	di	&	4.358	&	4.352	&	4.376	&	4.355	&	4.395	&	4.380	
&&	2.42	&	2.20	&	I	&	2.74	&	2.32	&	1.33	&	1.35	&	1.27	\\
BP	&	zb	&	4.538	&	4.527	&	4.547	&	4.510	&	4.566	&	4.540	
&&	2.40	&	2.20	&	I	&	2.43	&	2.12	&	1.28	&	1.41	&	1.24	\\
AlP	&	zb	&	5.463	&	5.459	&	5.481	&	5.438	&	5.498	&	5.472	
&&	2.50	&	2.45	&	I	&	2.71	&	2.44	&	1.53	&	1.83	&	1.56	\\
AlAs	&	zb	&	5.661	&	5.668	&	5.697	&	5.639	&	5.713	&	5.682	
&&	2.23	&	2.15	&	I	&	2.40	&	2.16	&	1.33	&	1.64	&	1.37	\\
GaP	&	zb	&	5.451	&	5.457	&	5.490	&	5.419	&	5.523	&	5.468	
&&	2.35	&	2.27	&	I	&	2.67	&	2.42	&	1.53	&	1.90	&	1.62	\\
GaAs	&	zb	&	5.648	&	5.668	&	5.711	&	5.626	&	5.745	&	5.687	
&&	1.52	&	1.42	&	D	&	1.86	&	1.18	&	0.43	&	0.52	&	0.42	\\
GaSb	&	zb	&	6.096	&	6.099	&	6.146	&	6.043	&	6.182	&	6.111	
&&	0.82	&	0.73	&	D	&	1.31	&	0.70	&	0.09	&	0.09	&	0.06	\\
InP	&	zb	&	5.869	&	5.872	&	5.914	&	5.840	&	5.961	&	5.891	
&&	1.42	&	1.34	&	D	&	2.23	&	1.61	&	0.83	&	0.91	&	0.83	\\
InAs	&	zb	&	6.058	&	6.075	&	6.125	&	6.038	&	6.172	&	6.099	
&&	0.42	&	0.35	&	D	&	0.93	&	0.36	&	0.00	&	0.00	&	0.00	\\
InSb	&	zb	&	6.479	&	6.486	&	6.540	&	6.430	&	6.588	&	6.501	
&&	0.24	&	0.18	&	D	&	0.80	&	0.28	&	0.00	&	0.00	&	0.00	\\
ZnS	&	zb	&	5.409	&	5.413	&	5.436	&	5.319	&	5.465	&	5.383	
&&	3.80	&	3.66	&	D	&	4.12	&	3.37	&	2.25	&	2.40	&	2.22	\\
ZnSe	&	zb	&	5.668	&	5.686	&	5.713	&	5.589	&	5.738	&	5.657	
&&	2.82	&	2.50	&	D	&	2.93	&	2.27	&	1.21	&	1.48	&	1.26	\\
ZnTe	&	zb	&	6.089	&	6.127	&	6.156	&	6.017	&	6.175	&	6.089	
&&	2.39	&	2.35	&	D	&	2.77	&	2.16	&	1.28	&	1.45	&	1.29	\\
CdSe	&	zb	&	6.052	&	6.118	&	6.156	&	6.024	&	6.195	&	6.098	
&&	1.90	&	1.74	&	D	&	1.90	&	1.36	&	0.34	&	0.73	&	0.45	\\
CdTe	&	zb	&	6.480	&	6.534	&	6.575	&	6.423	&	6.611	&	6.502	
&&	1.57	&	1.48	&	D	&	2.00	&	1.49	&	0.61	&	0.88	&	0.67	\\
$\beta$-GaN	&	wu	&	4.523	&	4.483	&	4.520	&	4.476	&	4.552	&	4.520	
&&	3.30	&	3.23	&	D	&	4.05	&	2.97	&	1.93	&	1.56	&	1.70	\\
$\alpha$-GaN	&	wu (a)	&	3.189	&	3.176	&	3.197	&	3.164	&	3.222	&	3.197	
			&&	3.50	&	3.44	&	D	&	4.23	&	3.14	&	2.08	&	1.74	&	1.85	\\
	&	wu (b)	&	5.185	&	5.167	&	5.215	&	5.171	&	5.246	&	5.216	
	&&	---	&	---	&	---	&	---	&	---	&	---	&	---	&	---	\\
InN	&	wu (a)	&	3.537	&	3.524	&	3.555	&	3.518	&	3.585	&	3.552	
			&&	0.72	&	0.69\cite{Arnaudov2004}	&	D	&	1.51	&	0.66	&	0.00	&	0.00	&	0.00	\\
	&	wu (c)	&	5.704	&	5.682	&	5.740	&	5.697	&	5.776	&	5.745	
	&&	---	&	---	&	---	&	---	&	---	&	---	&	---	&	---	\\
\hline
\multicolumn{10}{l}{\textbf{Group II: Wide-Gap/Notable}} &\\
C	&	di	&	3.567	&	3.536	&	3.556	&	3.538	&	3.580	&	3.563	
&&	5.50	&	5.48\cite{Yacobi2003}	&	I	&	6.11	&	5.42	&	4.19	&	4.19	&	4.05	\\
AlSb	&	zb	&	6.136	&	6.138	&	6.172	&	6.098	&	6.194	&	6.146	
&&	1.69	&	1.62	&	I	&	2.05	&	1.85	&	1.18	&	1.49	&	1.22	\\
BN	&	zb	&	3.616	&	3.586	&	3.605	&	3.585	&	3.629	&	3.611	
&&	6.36	&	6.20	&	I	&	6.69	&	5.90	&	4.40	&	4.47	&	4.30	\\
CdS	&	zb	&	5.818	&	5.865	&	5.899	&	5.776	&	5.944	&	5.844	
&&	2.55	&	2.31	&	D	&	2.72	&	2.10	&	1.01	&	1.34	&	1.08	\\
MgS	&	zb	&	5.622	&	5.670	&	5.690	&	5.619	&	5.518	&	5.675	
&&	4.78	&	---	&	D	&	5.17	&	4.48	&	3.37	&	3.68	&	3.34	\\
AlN	&	wu (a)	&	3.111	&	3.109	&	3.124	&	3.105	&	3.141	&	3.131	
			&&	3.60	&	3.50	&	I	&	6.62	&	5.50	&	4.29	&	3.96	&	3.97	\\
	&	wu (c)	&	4.981	&	4.967	&	4.999	&	4.968	&	5.030	&	5.009	
	&&	---	&	---	&	---	&	---	&	---	&	---	&	---	&	---	\\
MgTe	&	zb	&	6.420	&	6.445	&	6.469	&	6.362	&	6.504	&	6.432	
&&	6.19	&	6.13	&	D	&	3.91	&	3.49	&	2.57	&	2.96	&	2.58	\\
MgO	&	rs	&	4.207	&	4.200	&	4.222	&	4.178	&	4.249	&	4.231	
&&	7.90	&	7.30	&	D	&	7.87	&	6.40	&	4.90	&	4.58	&	4.50	\\
\hline
\multicolumn{10}{l}{\textbf{Group III: Benchmarking}\footnotemark[3] } &\\
BAs	&	zb	&	4.777	&	4.769	&	4.797	&	4.750	&	4.820	&	4.788	
&&	---	&	1.46	&	I	&	2.14	&	1.89	&	1.11	&	1.24	&	1.10	\\
MgSe	&	rs	&	5.460	&	5.484	&	5.508	&	5.418	&	5.518	&	5.476	
&&	---	&	2.47	&	I	&	3.05	&	2.58	&	1.62	&	2.01	&	1.69	\\
CaS	&	rs	&	5.689	&	5.696	&	5.704	&	5.582	&	5.710	&	5.640	
&&	---	&	---	&	I	&	4.02	&	3.52	&	2.10	&	2.62	&	2.27	\\
CaSe	&	rs	&	5.916	&	5.946	&	5.954	&	5.801	&	5.954	&	5.879	
&&	---	&	---	&	I	&	3.39	&	2.97	&	1.68	&	2.17	&	1.85	\\
CaTe	&	rs	&	6.348	&	6.385	&	6.390	&	6.211	&	6.388	&	6.294	
&&	---	&	---	&	I	&	2.66	&	2.32	&	1.19	&	1.67	&	1.35	\\
SrS	&	rs	&	5.990	&	6.030	&	6.043	&	5.924	&	6.055	&	5.985	
&&	---	&	4.10	&	I	&	3.97	&	3.52	&	2.20	&	2.73	&	2.38	\\
SrSe	&	rs	&	6.234	&	6.277	&	6.289	&	6.159	&	6.290	&	6.218	
&&	---	&	---	&	I	&	3.41	&	3.03	&	1.85	&	2.31	&	1.99	\\
SrTe	&	rs	&	6.640	&	6.698	&	6.707	&	6.539	&	6.706	&	6.615	
&&	---	&	---	&	I	&	2.82	&	2.51	&	1.42	&	1.93	&	1.59	\\
BaS	&	rs	&	6.389	&	6.401	&	6.414	&	6.298	&	6.433	&	6.350	
&&	---	&	3.88	&	I	&	3.61	&	3.21	&	1.99	&	2.55	&	2.15	\\
BaSe	&	rs	&	6.595	&	6.640	&	6.653	&	6.515	&	6.658	&	6.577	
&&	---	&	3.58	&	I	&	3.14	&	2.80	&	1.69	&	2.18	&	1.83	\\
BaTe	&	rs	&	7.007	&	7.002	&	7.018	&	6.854	&	7.024	&	6.919	
&&	---	&	3.08	&	I	&	2.48	&	2.22	&	1.25	&	1.73	&	1.38	\\
BSb	&	zb	&	5.278	&	5.222	&	5.256	&	5.201	&	5.280	&	5.240	
&&	---	&	---	&	D	&	1.48	&	1.31	&	0.74	&	0.76	&	0.70	\\
%------------------------------------------------------------------------------------------------------------------------------------------------------------------------	
\end{tabular}
\end{ruledtabular}
\footnotetext[1]{The abbreviations di, zb, wu and rs refer to the diamond, zincblende, wurtzite and rock salt lattices, respectively. The wurtzite structure is hexagonal, so there are two lattice vectors \emph{a} and \emph{c}. }
\footnotetext[2]{The abbreviations I or D refer to \emph{indirect} or \emph{direct} gap transitions; all experimental data taken from Reference~\onlinecite{Madelung2004} and references therein, unless otherwise noted. Low temperature experimental band gaps vary in terms of temperature, and are not available for all systems: \textbf{0 K (extrapolated)}: Si, GaP, GaAs, GaSb, CdTe\cite{vanDoorn1956}, MgS\cite{Davidson2010}, MgO; 
$\mathbf{\leq 2}$\textbf{K:} Ge, $\beta$-SiC, AlP, InP, InSb, ZnTe, $\alpha-$GaN, MgTe\cite{Hartmann1996};  
$\mathbf{\sim 4-6}$\textbf{K:} AlAs, InN,\cite{Anderson2006} InAs,  ZnSe; 
$\mathbf{\sim 7-10}$K: $\beta$-GaN, BN,\cite{Evans2008} AlN;
\textbf{14 K:} ZnS\cite{Berger1997} ;  \textbf{27 K:} AlSb ;  \textbf{Others:} CdSe,\cite{Zakharov1994} CdS,\cite{Zakharov1994} BAs. }
\footnotetext[3]{All experimental data from Table I, Reference~\onlinecite{Heyd2005} except for MgSe (Reference~\onlinecite{Madelung2004}.)}
\end{table*} % TTTTTTTTTTTTTTTTTTTTTTTTTTTTTTTTTTTTTTTTTTTTTTTTTTTTTTTTTTTTTTTTTTTTTTTTTTTTTTTTTTTTTTT
%

%|||||||||||||||||||||||||||||||||||||||||||
\subsection{Lattice Parameters} \label{sec:LatParams} % ======================================
%|||||||||||||||||||||||||||||||||||||||||||
 The LSDA provides reasonable, but generally contracted lattices, while gradient-corrected functionals tend to over-correct and expand them, instigating the development of functionals specifically-tailored to produce reasonable solid state geometries.
Arguably, the most well-known of these special functionals is PBEsol,\cite{PBEsol} a modified GGA with a significantly improved lattice constants for solids ($\sim$4x better) relative to PBE.
Nevertheless, the virtues of PBEsol were not seen to extend to semiconductors.\cite{PBEsol2} 
Interestingly, a recent assessment of meta-GGAs,\cite{Sun2011} for twenty solids utilizing experimental lattice parameters extrapolated to 0 K and then corrected with the Zero Point Anharmonicity Expansion (ZPAE),  indicates that LSDA does not perform as well as any of the special functionals studied.
Nevertheless, markedly different trends in performance are seen upon examination of a subset including only the five semiconductors and MgO present in the aforementioned test set.
A comparison of these errors, relative to corrected low temperature experiment, for LSDA, PBEsol, AM05,\cite{AM05} and revTPPS,\cite{revTPSS}  is extended to include the screened exchange hybrids HSE and HISS, as is summarized in Table~\ref{tab:sols}. 
%
%: TABLE Sol vs LSDA vs. HISS  TTTTTTTTTTTTTTTTTTTTTTTTTTTTTTTTTTTTTTTTTTTTTTTTT
\begin{table*}[!hb]\caption{\label{tab:sols} Comparison of equilibrium lattice parameters for five semiconductors and MgO relative to uncorrected\cite{TPSS} and corrected extrapolations to 0 K from experimental values, as provided in Ref.~\onlinecite{Sun2011}.  Mean error (ME) and mean average errors (MAE) reported in pm. (Corrected data is obtained via subtraction of the zero-point anharmonic expansion (ZPAE) from the extrapolated 0 K values. HISS and HSE values are from this work, the remainder are from Ref.~\onlinecite{Sun2011}.) }
\begin{ruledtabular}
\begin{tabular}{l cc cc cc cc}
         & \multicolumn{6}{c}{Theoretical $a_{0}$ (\AA)} &\multicolumn{2}{c}{Experimental $a_{0}$ (\AA)}\\
 \cline{2-7} \cline{8-9} \noalign{\smallskip}      
Solid & LSDA & PBEsol & AM05 & revTPSS & HSE & HISS & \textbf{0 K-ZPAE}& \textbf{0 K} \\
\hline
Ge & 5.631 & 5.680 & 5.685 & 5.682 & 5.706 & 5.660 & 5.644 & 5.658\\
Si & 5.405 & 5.432 & 5.434 & 5.439 & 5.451 & 5.430 & 5.421 & 5.430\\
$\beta$-SiC & 4.332 & 4.355 & 4.352 & 4.357 & 4.376 &4.352 & 4.346 & 4.358\\
GaAs & 5.615 & 5.665 & 5.672 & 5.680 & 5.711 & 5.668 & 5.640 & 5.648\\
C & 3.533 & 3.552 & 3.548 & 3.558 & 3.556 & 3.535 & 3.553 & 3.567\\
MgO & 4.170 & 4.223 & 4.229 & 4.240 & 4.222 & 4.200& 4.189 & 4.207\\
\hline
ME&	-2.5	& 1.9 & 2.1 &2.7 &3.8 & 0.9	 & \multicolumn{2}{l}{ZPAE-corrected}  \\
MAE &1.8 & 1.9& 2.3 & 2.7 & 3.8 & 1.4	 &\multicolumn{2}{l}{extrapolation to 0 K} \\
\hline  
ME  & -3.0 &	0.7 &	0.9 &	1.5 &	2.6 &	-0.4 & \multicolumn{2}{l}{Uncorrected} \\
MAE & 3.0 &	1.3 &	1.7 &	1.8 &	2.9 &	1.1 & \multicolumn{2}{l}{extrapolation to 0 K} \\	
%-------------------------------------------------------------------------------------------------------------------------------------------------------------	
\end{tabular}
\end{ruledtabular}
\end{table*} % TTTTTTTTTTTTTTTTTTTTTTTTTTTTTTTTTTTTTTTTTTTTTTTTTTTTTTTTTTTTTTTTTTTTTTTTTTTTT

For this test set, the mean error (ME) of LSDA is actually larger than that of PBEsol for the corrected low temperature lattice parameters, and of similar magnitude (but opposite sign) relative to revTPSS.
HSE has the largest ME, but HISS -- not LSDA -- provides the best lattice parameters.
Considering the mean absolute error (MAE) instead, LSDA is only 0.1 pm better than PBEsol, while HISS again proves superior to both. 
Note also that the middle-range screened hybrid yields better lattice parameters for the narrow- and wide-gap cubic semiconductors, as well the ionic MgO, even for the uncorrected extrapolations to absolute zero.

Admittedly, this test set of six solids is rather small, and while the earlier semiconductor test set utilized for HISS\cite{HISS1} showed promising results for a slightly different set from that contained in Table~\ref{tab:sols}, it is also small.
We thus examined the performance of HISS for the entire SC40 test set and compared the data to that produced by the LSDA, TPSS, PBEsol, TPSS, and HSE functionals.
This provides a much larger \textit{semiconductor} test set while allowing for analysis of some wider-gap materials commonly used in conjunction with the most commonly used materials in Group 1 of Table~\ref{tab:sc40}.
Given that the PBEsol lattice constants are of higher quality than PBE, we do not include data for the latter in this table for aesthetic reasons. 
Nevertheless, since both GGAs yield band gaps of similar magnitude, PBE data is presented in other analyses, particularly those in Section~\ref{subsec:Gaps}.

\begin{figure}[!hb]
	\includegraphics[width=3.375in]{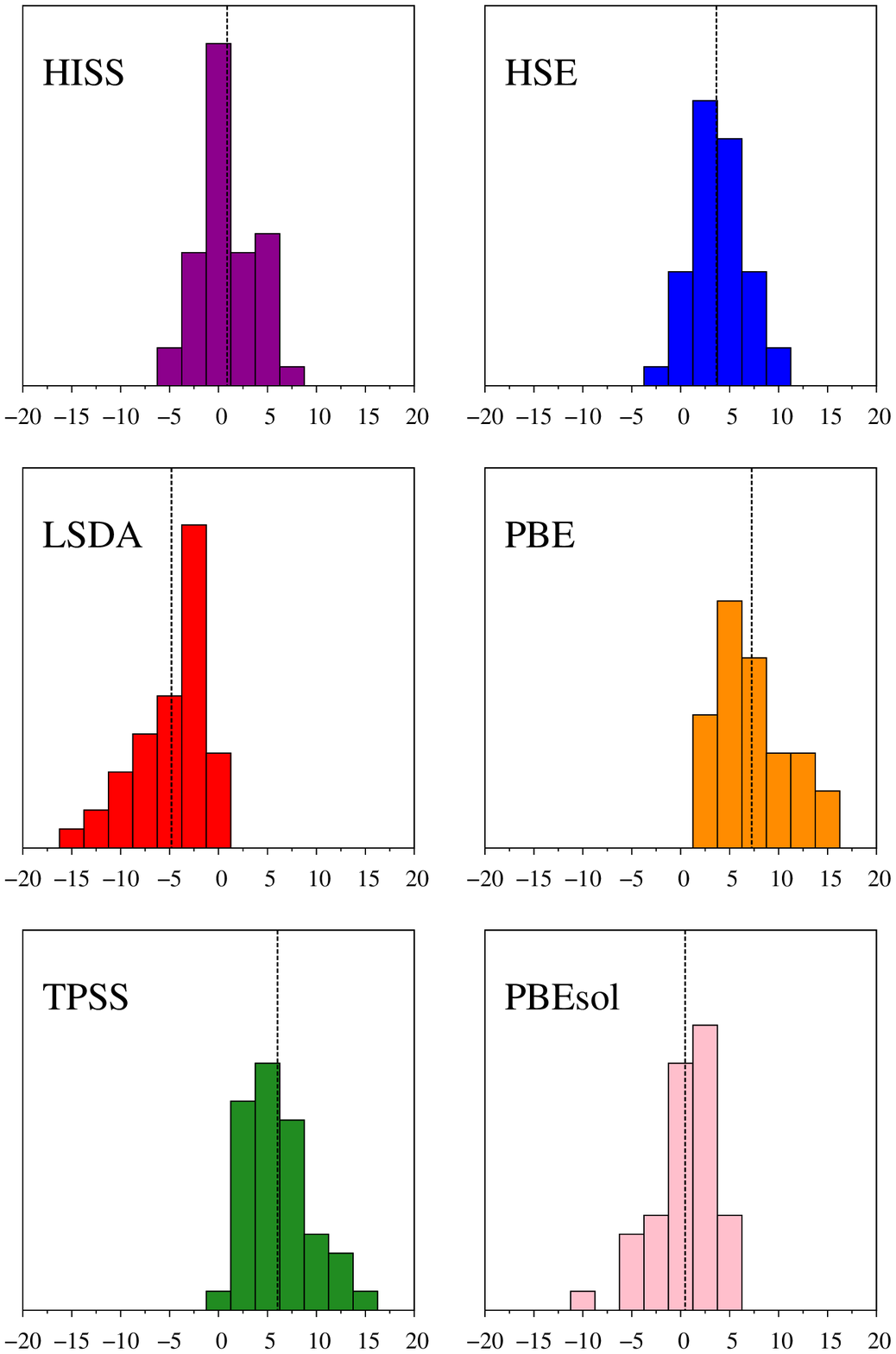}
	\caption{\label{fig:histograms} Histograms of calculated lattice constants errors (pm) relative to experimental values from Table~\ref{tab:sc40}. The dashed lines indicate MEs for HISS, HSE, LSDA, PBE, TPSS and PBEsol: 0.9, 3.7, -4.8, 7.3, 6.0 and 0.5 pm, respectively.}
\end{figure}

 To better illustrate the magnitudes and systematics of lattice parameter
  predictions, Figure~\ref{fig:histograms} presents the errors for all forty-three lattice
  parameters contained in Table~\ref{tab:sc40}, as well as those for PBE, in the form
  of histograms. 
  The ME for each functional is depicted by the vertical dotted line on each plot.
%
%: Figure Six LP Histograms
The results from HISS are sharply peaked around zero error, it is clear that the LSDA underestimates lattice constants, the well-known PBE and TPSS overestimations are evident, and HSE produces highly-clustered overestimations with accuracy comparable to that of the LSDA. 
The PBEsol distribution is somewhat narrower than that of LSDA, but, as with other GGAs, is slightly more skewed toward expanded lattice constants, with the notable exception of the Ca, Sr and Ba chalcogenides. 
PBEsol actually contracts these Group 3 lattices and performs more like LSDA, particularly in the case of BaTe, the obvious outlier in Figure~\ref{fig:histograms}. 
%
%: TABLE Errors Gaps  1 & 2 only   TTTTTTTTTTTTTTTTTTTTTTTTTTTTTTTTTTTTTTTTTTTTTTTTTTTTTTTTTTTTT
\begin{table*}\caption{\label{tab:Errors2} Error Statistics  for equilbrium geometries (pm) and band gaps (eV) calculated at those geometries. The band gap data is compared relative to low (LT) and room (RT) temperature experiment. Groups 1-3 defined in Table~\ref{tab:sc40}. }
\begin{ruledtabular}
\begin{tabular}{ll ccccc rr  rrrrrr}
&& \multicolumn{5}{c}{\textbf{Lattice Parameters} (pm)} & &\multicolumn{5}{c}{\textbf{Band Gaps} (eV)}\\
 \cline{3-7} \cline{9-13}\noalign{\smallskip}\\
 &Error\footnotemark[1]  & HISS & HSE & LSDA &TPSS & PBEsol &&HISS & HSE & LSDA & TPSS & PBEsol  \\
\hline
Group 1 && & & & & &LT Group 1  &\\
%      & HISS & HSE & LSDA & TPSS & PBEsol        & HISS & HSE & LSDA & TPSS & PBEsol
&ME		& 0.53 & 4.09 & -3.41 & 7.14 & 1.72 &  ME	& 0.39 &-0.17 & -0.98 & -0.88 & -1.00 \\
&MAE	& 1.69 & 4.12 &  3.41 & 7.14 & 2.09 & MAE	& 0.39 & 0.21 & 0.98 & 0.88 & 1.00\\
Groups 1+2  && & & & && 	LT Groups 1+2  &\\
&ME		& 0.50 & 3.74 & -3.22 & 6.71 & 1.75 & ME	& 0.38 &-0.24 & -1.15 & -0.93 & -1.19 \\
&MAE	& 1.87 & 3.90 & 3.22 & 6.71 & 2.07 & MAE	& 0.38 & 0.28 & 1.15  & 0.93  & 1.19\\
Group 3 & & & & & &&	RT Group 1 &\\
& ME	& 1.88 & 3.42 &-8.93 & 4.27 & -2.85 & ME	& 0.50 & -0.07 &-0.88 & -0.78 & -0.89\\
&MAE	& 3.05 & 3.79  & 8.93 & 4.27 & 3.30 &MAE	& 0.50 & 0.15 & 0.88 & 0.78  & 0.89\\
All Groups & & & & & &&	RT Groups 1+2 &\\ 
& ME	& 0.89 & 3.65 & -4.81 & 6.03 & 0.46 & ME	& 0.49 & -0.12 &-1.02 &-0.93  & -1.06\\
&MAE	& 2.20 & 3.87  & 4.81 & 6.03 & 2.42 &MAE	& 0.49 & 0.20 & 1.02 &  0.93 & 1.06\\
\end{tabular}
\end{ruledtabular}
\footnotetext[1]{MAE=Mean Absolute Error; Max=Maximum absolute error}
\end{table*}
% TTTTTTTTTTTTTTTTTTTTTTTTTTTTTTTTTTTTTTTTTTTTTTTTTTTTTTTTTTTTTTTTTTTTTTTTTTTTTTTTTTTTTTTTTT
%

\begin{figure}[!htb]
	\includegraphics[width=3.375in]{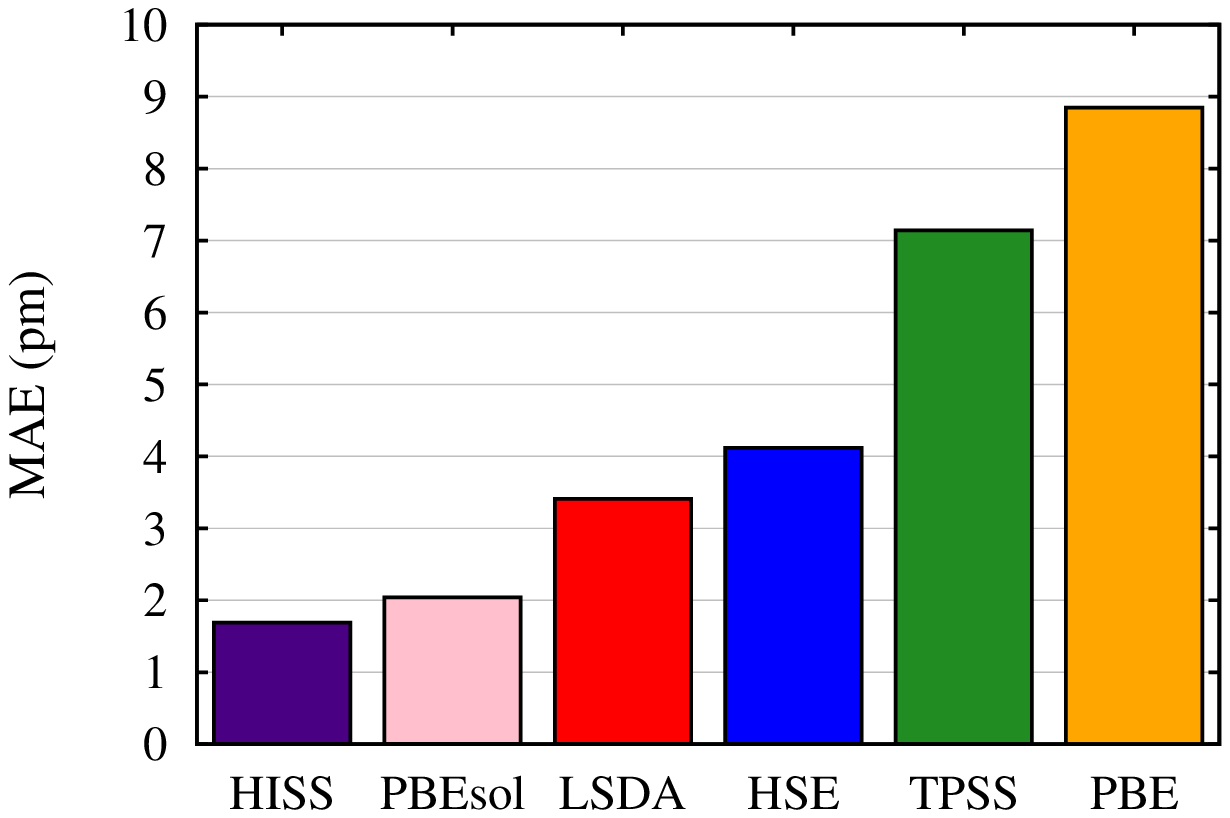} 
	\caption{\label{fig:LPs} MAEs comparing equilibrium/relaxed lattice parameters to the experimental data listed in Group 1 of Table~\ref{tab:sc40}. This data is summarized in Table~\ref{tab:Errors2}. Notice the exceptional performance of HISS.}  
	\end{figure}

As was demonstrated from the smaller test set in Table~\ref{tab:sols}, it is helpful to consider both ME and MAE in performance evaluations.
The mean absolute errors of the equilibrium lattice parameters for the twenty commercially-relevant semiconductors in Group 1 are thus presented in Figure \ref{fig:LPs}.
%
%:  figure Group 1 Lattice Params
% figure  

For this particular group, PBEsol \emph{does} outperform LSDA, but not HISS. 
Remarkably, HISS provides excellent lattice parameters, with a MAE of less than half that of LSDA, (see Table~\ref{tab:Errors2}) despite producing bond lengths in molecules that are not particularly impressive.\cite{HISS2}
In fact, when comparing the ten largest deviations from experiment, HISS outperforms all semilocal functionals other than PBEsol for the zinc and cadmium chalcogenides, produces more accurate lattice parameters for the $\alpha$-GaN and InN wurtzite structures and excels for the diamond lattices.
HISS, like PBEsol also performs well for Groups 2 and 3, while the LSDA errors are notably larger for Group 3.
All of the error data is summarized by group in Table~\ref{tab:Errors2}.

%|||||||||||||||||||||||||||||||||||||||||||
\subsection{Band Gaps} \label{subsec:Gaps} % ====================================================
%|||||||||||||||||||||||||||||||||||||||||||

\begin{figure}[!htb]
	\includegraphics[width=3.375in]{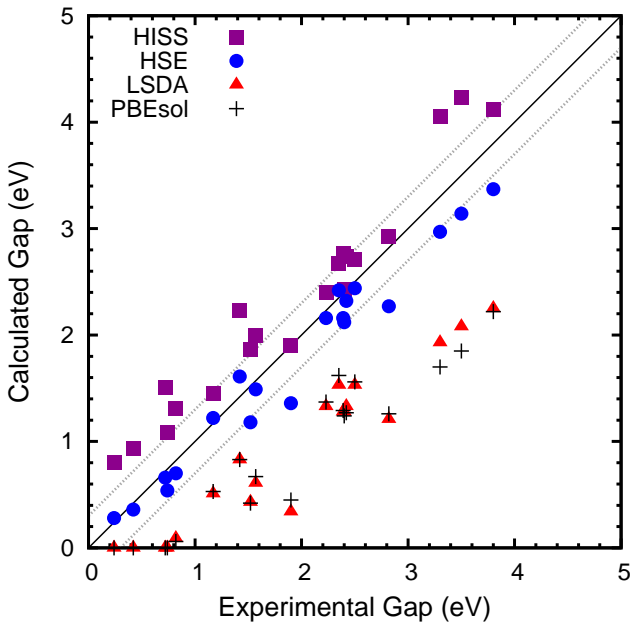} 
	\caption{\label{fig:scattergaps} Comparison of DFT band gaps vs. low temperature experimental gaps, for Group I of Table~\ref{tab:sc40}.  The 20 semiconductors have gaps in the visible range. Dotted lines indicate deviations of $\pm$ 0.3 eV. }  
	\end{figure}

It should be emphasized that screened hybrids, such as HISS and HSE, which calculate nonlocal Hartree-Fock type exchange with the GKS approach, predict fundamental gaps as band energy (conduction band - valence band) differences.\cite{MCY2009}
They also provide excellent approximations of optical gaps, particularly for visible-range semiconductors and other materials.\cite{Brothers2008}
This is well-illustrated in  Figure~\ref{fig:scattergaps}, which plots the mean absolute errors (MAEs) for the calculated HISS, HSE, LSDA and PBEsol band gaps of Group 1 (Table~\ref{tab:sc40}) against low temperature experimental gaps.
The dashed lines indicate deviations of $\pm 0.3$ eV.
%
%: figure TheHiss20 
% figure  
%

Almost all HSE data is within the $\pm 0.3$ eV range, while HISS tends to run toward larger gaps and the semilocal functionals consistently underestimate all band gaps by more than 0.3 eV.
Our group previously demonstrated that HSE is capable of correctly describing the Mott metal-insulator transition in actinide oxides: materials known to be narrow-gap semiconductors.\cite{Prodan2007,Prodan2006,Prodan2005} 
All of the semilocal functionals predict these systems to be metallic.
Similarly, for the SC40 test set, the LSDA metallizes the narrow band gaps of Ge and the indium pnictides as is evident from he clustering on the x-axis between 0 and 1 in  Figure~\ref{fig:scattergaps}.  
HSE -- and HISS -- not only predict narrow gaps, but do so in good agreement with experiment.

It is interesting to note that while PBEsol and HSE yield similar lattice parameters for nitride compounds adopting the wurtzite structure, the band gaps are significantly different: HSE band gaps are in much better agreement with experiment. 
As is evident from Table~\ref{tab:sc40}, HISS consistently produces better lattice parameters than semilocal functionals accompanied by systematically-larger band gaps.
These overestimations are, nevertheless, typically smaller in magnitude than the well-known underestimations expected from semilocal functionals.

\begin{figure}[!htb]
	\includegraphics[width=3.375in]{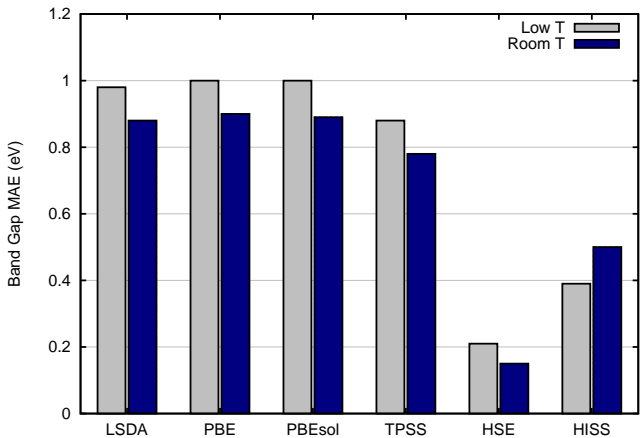} 
	\caption{\label{fig:MAEgrp1}Mean absolute errors relative to both low (gray) and room (blue) temperature experimental band gaps for the Group 1semiconductors in Table~\ref{tab:sc40}. The 20 semiconductors absorb in the visible range and adopt  diamond, zincblende and wurtzite structures.}  
	\end{figure}

The MAEs for Groups 1 and 2 are summarized on the right of Table~\ref{tab:Errors2}.
As expected, HSE shows the smallest errors, on the order of 0.20 eV for either temperature data set, but HISS deviates less for more ionic compounds like MgO or the Mg chalcogenides, and yields better lattice parameters.
The performance of all functionals for Group 1 semiconductors relative to both temperature ranges is summarized in in Figure~\ref{fig:MAEgrp1}.
In marked contrast to all other functionals, HISS appears to have errors in better agreement with respect to the reported LT gaps than the RT gaps, an artifact of the systematic overestimation of band gaps demonstrated in Figure~\ref{fig:scattergaps}.
%
%: figure New21band gaps all functions.
% figure  

%|||||||||||||||||||||||||||||||||||||||||||
\subsection{Noteworthy Cases} \label{sec:Ack} % ====================================================
%|||||||||||||||||||||||||||||||||||||||||||

Having discussed the bulk of the statistics above, we now turn our attention to a few systems worthy of comment.
As is pointed out in the Springer Handbook,\cite{Madelung2004} there is conflicting experimental data regarding the magnitude  and, at one time, some uncertainty regarding the location of the AlSb indirect transition.
One set of experiments indicates that the indirect transition occurs somewhere near $\Gamma \rightarrow$ X, with a magnitude of 1.6-1.7 eV,\cite{Alibert1983, Mathieu1975} while another set supports a $\Gamma \rightarrow$ L transition\cite{Joullie1982} with a gap closer to 2 eV, and in recent texts, the gap is reported to be ca. 1.5, 1.6 or 1.7 eV in Refs.~\onlinecite{Grundmann2010,Springer2006} and~\onlinecite{Economou2010}, respectively.

\begin{figure*}[!htbp]
\includegraphics[width=2.3in]{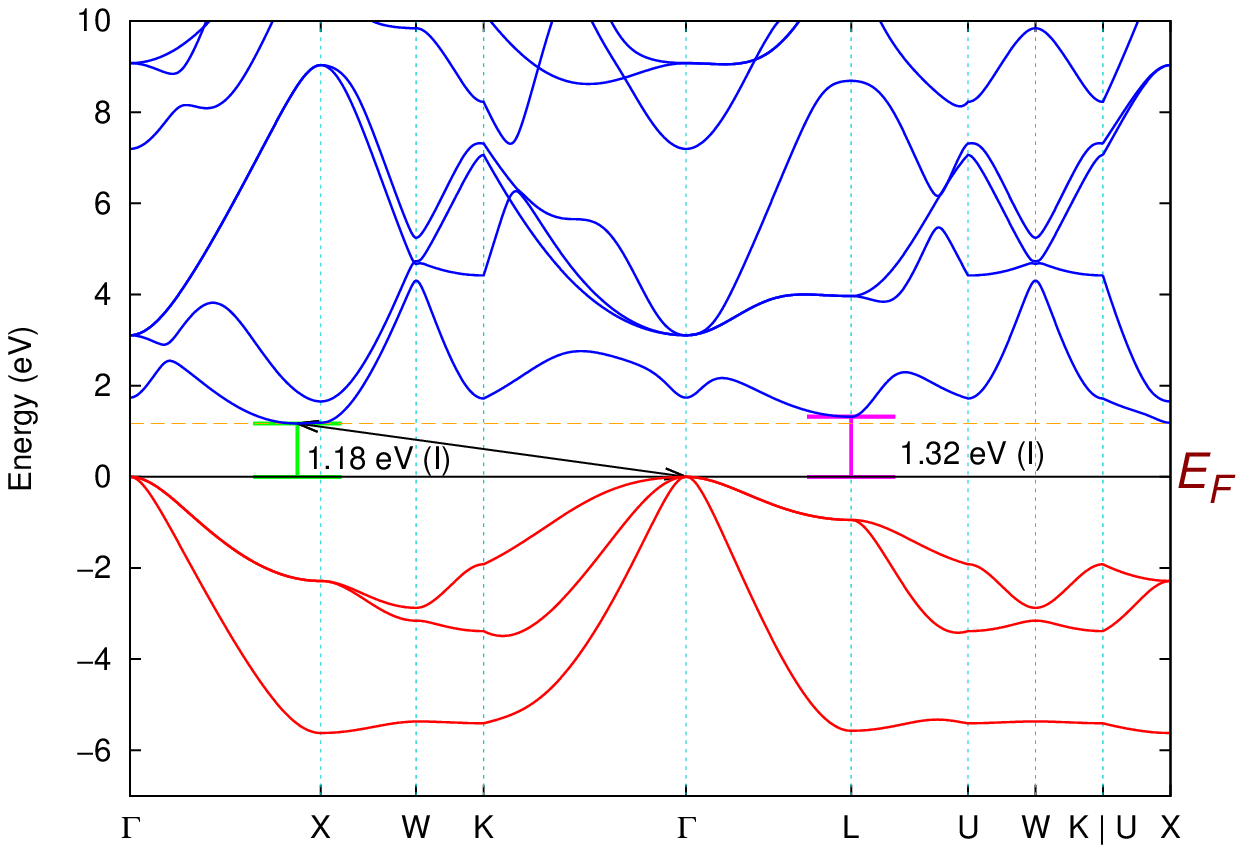}
\includegraphics[width=2.3in]{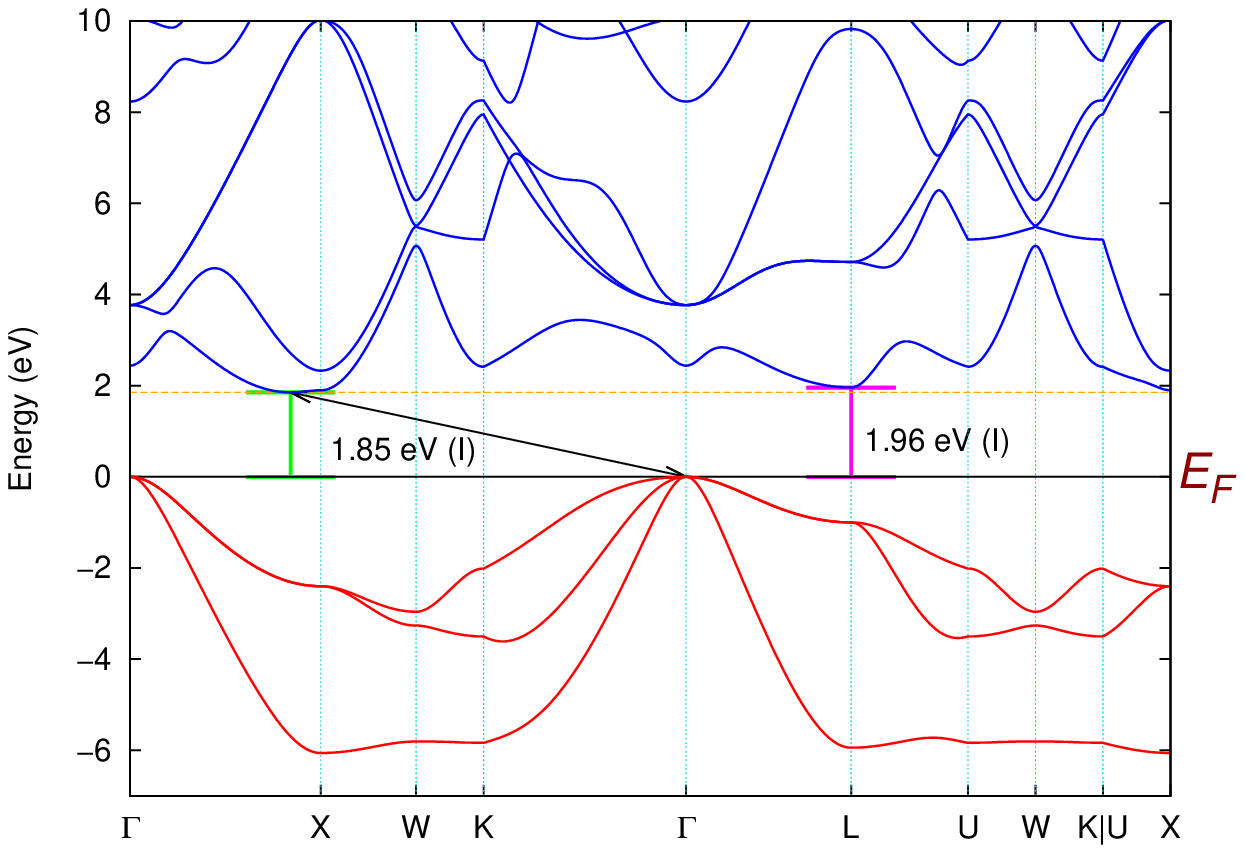}
\includegraphics[width=2.3in]{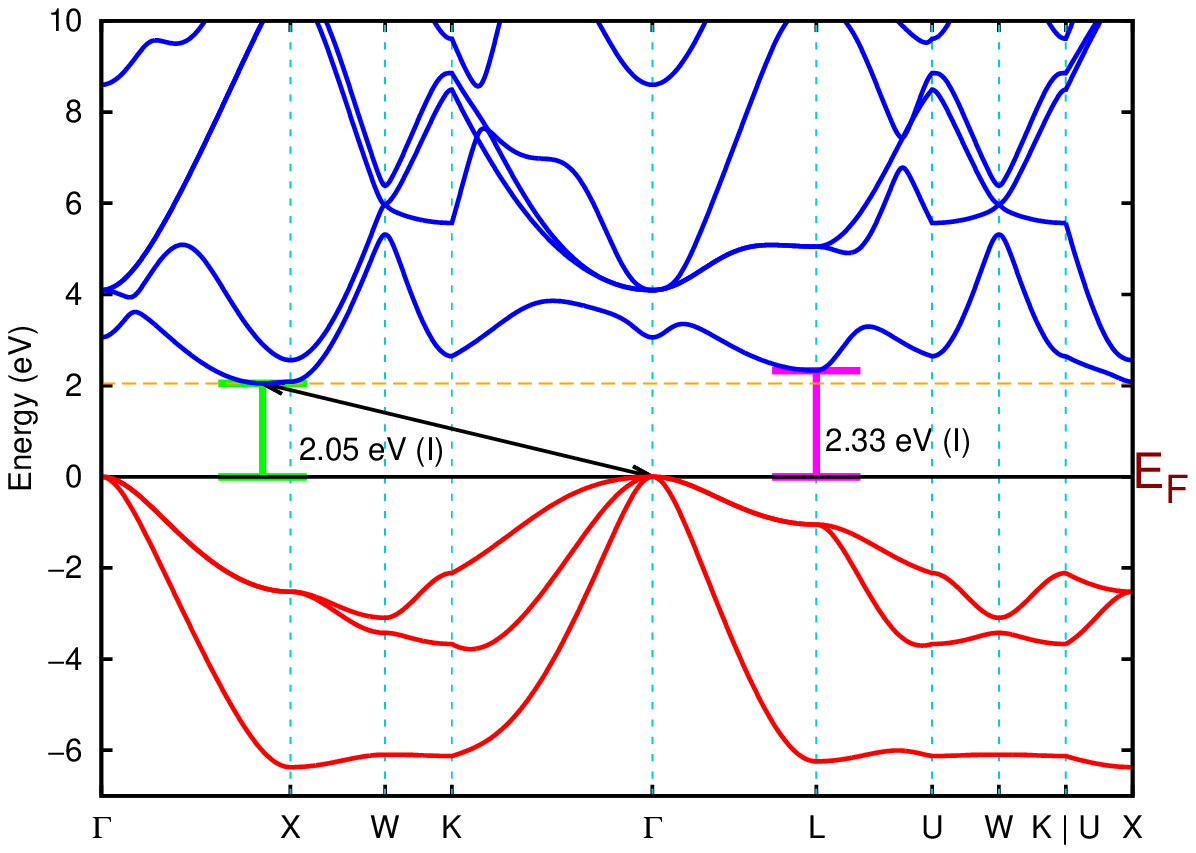}
\caption{\label{fig:AlSb} Band structures for the relaxed geometries of AlSb as calculated with LSDA (left), HSE (middle) and HISS (right). Note that all functionals show that the two observed experimental transitions near $\Gamma \to X$ and $\Gamma \to L$ are close in energy, but that the $\Gamma \to X$ gap is smaller.}
\end{figure*}

We calculated the band structure of AlSb at the equilibrium geometries produced by LSDA, HSE and HISS.
All functionals indicate that the indirect transition is clearly in the vicinity of near $\Gamma \rightarrow$ X, Figure~\ref{fig:AlSb}. 
The 0 K HSE gap of 1.82 eV compares favorably with the 1.69 eV (27 K) experimental band gap, while showing the characteristic camel's back structure.

The slightly larger 2.05 eV band gap predicted by HISS might encourage preference of the $\Gamma \rightarrow$ L transition, but as is seen from plot to the right of Figure~\ref{fig:AlSb}, the smallest indirect gap again occurs near X and there is little difference between the structure of highest valence and lowest conduction bands for either screened hybrid calculation.
The fact that HISS \textit{systematically} overestimates the band gaps, yet shows similar structure to HSE also tends to favor the smaller transition at X.
Note also that the width of the valence band predicted by HISS is intermediate between that of LSDA and HSE.

The  zincblende polymorph of CdS, Hawleyite, included in the SC40 test set is not as widely used as wurtzite CdS, Greenockite: the hexagaonal form is commonly employed in thin film solar cells and quantum dots. 
Note that most of the available, reliable experimental data for the cubic CdS is for thin films, which typically have larger band gaps than for the bulk, which is what we calculate here.
Nevertheless, the HISS and HSE results are within a reasonable range and the consistently lower errors HISS yields compared to those of semilocal functionals are encouraging.

Finally, the wide-gap nitrides of aluminum and boron are worthy of comment because BN is an unsual zincblende compound, while the isoelectronic AlN adopts a rather polar wurtzite structure.
Boron nitride, ``borazone'' is almost as hard as diamond, has excellent refractory properties,\cite{Berger1997} and, atypical of most zincblende structures, has a large band gap (6.3 eV), well into the UV spectrum.
Interestingly, the predicted screened hybrid band gaps for these systems (Table~\ref{tab:sc40}) remain in good agreement with experiment, implying some degree of utility for higher-energy semiconductor applications.

%|||||||||||||||||||||||||||||||||||||||||||
\subsection{Timings} \label{sec:Timings} % ====================================================
%|||||||||||||||||||||||||||||||||||||||||||
In addition to excellent performance for lattice parameters, HISS possesses an another advantage over HSE: because the range-separation parameter $\omega_{LR}$ for HISS is roughly twice the parameter $\omega$ of HSE, the long-range exchange in HISS is screened more aggressively.  
Accordingly, the computational cost for HISS is smaller than that for HSE.  
The claim in Reference~\onlinecite{Hirao2011} that HISS is quite expensive because one must do the two-electron integrals twice is simply incorrect.  
While this is true in the most naive implementation possible, computationally efficient implementations of HISS will calculate only the so-called $[0]^{(m)}$ integrals twice, and the expensive steps of contraction and angular moment incrementation are performed once, just as in HSE.

Table \ref{tab:Timings} contains the CPU time per SCF cycle relative to that of PBE for BP, C, InN, MgO, Si, and SiC.  
This set includes both wide gap and narrow gap materials adopting diamond, zincblende or wurtzite lattices.  
All calculations are performed at the experimental geometries.  
%
%: TABLE Relative Timings     TTTTTTTTTTTTTTTTTTTTTTTTTTTTTTTTTTTTTTTTTTTTTTTTTTTTTTTTTTTTT
\begin{table}[!htb]\caption{\label{tab:Timings} CPU time per SCF cycle relative to PBE for representative systems. The numbers in parentheses are the number of SCF cycles required for convergence.}
\begin{ruledtabular}
\begin{tabular}{l llll}
& \multicolumn{4}{c}{Relative Avg. Time (\# of SCF Cycles)}\\
	\cline{2-5}\noalign{\smallskip}
Solid	   &PBE & HISS & HSE06 & PBEh \\
\hline
C  & (9) & 1.95 (9)   & 2.63 (10) &  --- \\ 
Si  & (10)  & 1.70 (12)  & 2.38 (10)  & 33.18 (11)  \\
BP  &  (8) & 1.88 (9) & 2.56 (9) & 37.75 (9)\\  
SiC  &  (8) & 1.89 (8) & 2.64 (9) & 39.51 (9) \\
MgO & (10)  & 1.85 (10)  & 2.73 (8) & 32.77 (15) \\
InN  & (15) & 2.21 (9) & 2.91 (7) & --- \\
\hline
Mean & 1 & 1.91 & 2.64 & 35.50 \\
\end{tabular}
\end{ruledtabular}
\end{table}
% TTTTTTTTTTTTTTTTTTTTTTTTTTTTTTTTTTTTTTTTTTTTTTTTTTTTTTTTTTTTTTTTTTTTTTTTTTTTTTTTTTTTTTTTTT
%% 
It is clear that HISS is only twice as expensive as PBE, and that HSE costs about 50\% more than HISS, both producing much better band gaps and lattice parameters.
We also include results for the PBEh\cite{PBE0} global hybrid (also known by the acronyms PBE0 and PBE1PBE) to illustrate the enormous savings resulting from range separation.
PBEh calculations on C and InN were not possible, nevertheless, the PBEh band gaps for Si, SiC BP, and MgO are respectively, 1.81, 3.22,  2.74, and 7.31 eV.
The PBEh band gaps are in error by approximately 0.7 eV, offering minimal improvement over PBE but at an enormously larger (40\by) computational cost.

Global hybrid functionals like B3LYP and PBEh have been benchmarked against HSE for a few systems and properties, and it has been shown that their increased computational cost (observed for both with plane waves and gaussian orbitals) does not lead to improved answers compared to HSE. 
This is even more true for semiconductors: both B3LYP and PBEh overestimate the band gaps\cite{MPSK2008}

%|||||||||||||||||||||||||||||||||||||||||||
\section{Concluding Remarks} \label{sec:Discussion} %--------------------------------------------------------------------------------
%|||||||||||||||||||||||||||||||||||||||||||

The middle-range screening utilized in the HISS functional allows for accurate reproduction of lattice parameters with concomitant band gap prediction at nearly the quality of HSE. 
The HISS screened hybrid band gap errors are systematically over-estimated, but are of increased accuracy for larger gap systems, while lattice parameters are superior to those produced by LSDA -- or even PBEsol -- for commercially-relevant semiconductors, including non-cubic systems.  

In fact, as was recently demonstrated for the perovskite SrTiO$_{3}$ system,\cite{Fadwa2011} HISS can rapidly reproduce difficult lattice distortions while producing near-HSE-quality band gaps.
The predictive power provided by such systematic errors was also nicely demonstrated by the disputation of the experimental geometry for the novel Delafossite material CuBO$_{2}$,\cite{SnureCuBO2,ScanlonCuBO2,TraniCuBO2} where both the accuracy of the band gaps and the generally systematic deviations in lattice parameters produced by HSE were used to justify theoretical arguments.
Given that HISS performs well for molecules,\cite{HISS1} it is reasonable to assume that molecule-surface interactions will show an improvement over what is currently available.

Finally, timing data indicates that HISS is more computationally efficient than HSE, yet more accurate than semilocal functionals, PBEsol or PBEh,  imbuing HISS with yet another advantage: reliable theoretical data for semiconductors may be acquired by running a calculation that will complete at a computational expense of merely twice that of PBE, without necessitating additional high-level corrections.
Such easily-obtained, high quality lattice constants coupled with improved optical band gap predictions will render HISS a useful adjunct to HSE in theoretical investigations of geometry-sensitive semiconductors and many wider-gap materials requiring accurate lattice parameters in order to obtain reliable solid state properties.

% \\\\\\\\\\\\\\\\\\\\\\\\\\\\\\\\\\\\\\\\\\\////////////////////////////////////////////////////////
%: ||||||||||||||||||||||||||||||||||||   ACKNOWLEDGMENTS    |||||||||||||||||||||||||||||||||||||||||||||||||||||||| 
\begin{acknowledgments}
The work at Rice University was supported by the Qatar National Research Fund (QNRF) through the National Priorities Research Program (NPRP Grant No.~09-431-076) and The Welch Foundation (C-0036). The authors would like to thank Dr. Donghyung Lee and Mr. Ireneusz W. Bulik helpful discussions and technical assistance.\\

\end{acknowledgments}

% ||||||||||||||||||||||||||||||||||||||||||||||||||||||||||||||||||||||||||||||||||||||||||||||||||||||||||||||||||||||||||||||||||||||||||
%: |||||||||||||||||||||||||||||||||||||    BIBLIOGRAPHY    ||||||||||||||||||||||||||||||||||||||||||||||||||||||||||   
% ||||||||||||||||||||||||||||||||||||||||||||||||||||||||||||||||||||||||||||||||||||||||||||||||||||||||||||||||||||||||||||||||||||||||||

%% <<<<<<<<<<<<<<<<<<    END OF FILE HISS-Oct23.tex   >>>>>>>>>>>>>>>>>>>>>>
\bibstyle{unsrt}
\bibliography{HISS}

%merlin.mbs 2010-03-15 4.21a (PWD, AO, DPC)
\begin{thebibliography}{89}%
\makeatletter
\providecommand \@ifxundefined [1]{%
 \@ifx{#1\undefined}
}%
\providecommand \@ifnum [1]{%
 \ifnum #1\expandafter \@firstoftwo
 \else \expandafter \@secondoftwo
 \fi
}%
\providecommand \@ifx [1]{%
 \ifx #1\expandafter \@firstoftwo
 \else \expandafter \@secondoftwo
 \fi
}%
\providecommand \natexlab [1]{#1}%
\providecommand \enquote  [1]{``#1''}%
\providecommand \bibnamefont  [1]{#1}%
\providecommand \bibfnamefont [1]{#1}%
\providecommand \citenamefont [1]{#1}%
\providecommand \href@noop [0]{\@secondoftwo}%
\providecommand \href [0]{\begingroup \@sanitize@url \@href}%
\providecommand \@href[1]{\@@startlink{#1}\@@href}%
\providecommand \@@href[1]{\endgroup#1\@@endlink}%
\providecommand \@sanitize@url [0]{\catcode `\\12\catcode `\$12\catcode
  `\&12\catcode `\#12\catcode `\^12\catcode `\_12\catcode `\%12\relax}%
\providecommand \@@startlink[1]{}%
\providecommand \@@endlink[0]{}%
\providecommand \url  [0]{\begingroup\@sanitize@url \@url }%
\providecommand \@url [1]{\endgroup\@href {#1}{\urlprefix }}%
\providecommand \urlprefix  [0]{URL }%
\providecommand \Eprint [0]{\href }%
\@ifxundefined \urlstyle {%
  \providecommand \doi  [0]{\begingroup \@sanitize@url \@doi}%
  \providecommand \@doi [1]{\endgroup \@@startlink {\doibase
  #1}doi:\discretionary {}{}{}#1\@@endlink }%
}{%
  \providecommand \doi  [0]{doi:\discretionary{}{}{}\begingroup
  \urlstyle{rm}\Url }%
}%
\providecommand \doibase [0]{http://dx.doi.org/}%
\providecommand \Doi [0]{\begingroup \@sanitize@url \@Doi }%
\providecommand \@Doi  [1]{\endgroup\@@startlink{\doibase#1}\@@Doi}%
\providecommand \@@Doi [1]{#1\@@endlink}%
\providecommand \selectlanguage [0]{\@gobble}%
\providecommand \bibinfo  [0]{\@secondoftwo}%
\providecommand \bibfield  [0]{\@secondoftwo}%
\providecommand \translation [1]{[#1]}%
\providecommand \BibitemOpen [0]{}%
\providecommand \bibitemStop [0]{}%
\providecommand \bibitemNoStop [0]{.\EOS\space}%
\providecommand \EOS [0]{\spacefactor3000\relax}%
\providecommand \BibitemShut  [1]{\csname bibitem#1\endcsname}%
%</preamble>
\bibitem [{\citenamefont {Scuseria}\ and\ \citenamefont
  {Staroverov}(2005)}]{Staroverov2005}%
  \BibitemOpen
  \bibfield  {author} {\bibinfo {author} {\bibfnamefont {G.~E.}\ \bibnamefont
  {Scuseria}}\ and\ \bibinfo {author} {\bibfnamefont {V.~N.}\ \bibnamefont
  {Staroverov}},\ }in\ \href@noop {} {\emph {\bibinfo {booktitle} {Theory and
  Applications of Computational Chemistry: The First 40 Years}}},\ \bibinfo
  {editor} {edited by\ \bibinfo {editor} {\bibfnamefont {C.~E.}\ \bibnamefont
  {Dykstra}}, \bibinfo {editor} {\bibfnamefont {G.}~\bibnamefont {Frenking}},
  \bibinfo {editor} {\bibfnamefont {K.~S.}\ \bibnamefont {Kim}}, \ and\
  \bibinfo {editor} {\bibfnamefont {G.~E.}\ \bibnamefont {Scuseria}}}\
  (\bibinfo  {publisher} {Elsevier},\ \bibinfo {address} {Amsterdam, The
  Netherlands},\ \bibinfo {year} {2005})\ pp.\ \bibinfo {pages}
  {669--724}\BibitemShut {NoStop}%
\bibitem [{\citenamefont {Dreuw}\ and\ \citenamefont
  {Head-Gordon}(2006)}]{Dreuw2006}%
  \BibitemOpen
  \bibfield  {author} {\bibinfo {author} {\bibfnamefont {A.}~\bibnamefont
  {Dreuw}}\ and\ \bibinfo {author} {\bibfnamefont {M.}~\bibnamefont
  {Head-Gordon}},\ }\href@noop {} {\bibfield  {journal} {\bibinfo  {journal}
  {Chem. Rev.},\ }\textbf {\bibinfo {volume} {105}},\ \bibinfo {pages} {4009}
  (\bibinfo {year} {2006})}\BibitemShut {NoStop}%
\bibitem [{\citenamefont {Zhao}\ and\ \citenamefont
  {Truhlar}(2006)}]{Zhao2006}%
  \BibitemOpen
  \bibfield  {author} {\bibinfo {author} {\bibfnamefont {Y.}~\bibnamefont
  {Zhao}}\ and\ \bibinfo {author} {\bibfnamefont {D.~G.}\ \bibnamefont
  {Truhlar}},\ }\href@noop {} {\bibfield  {journal} {\bibinfo  {journal} {J.
  Phys. Chem. A},\ }\textbf {\bibinfo {volume} {110}},\ \bibinfo {pages}
  {13126} (\bibinfo {year} {2006})}\BibitemShut {NoStop}%
\bibitem [{\citenamefont {Peach}\ \emph {et~al.}(2008)\citenamefont {Peach},
  \citenamefont {Benfield}, \citenamefont {Helgaker},\ and\ \citenamefont
  {Tozer}}]{Peach2008}%
  \BibitemOpen
  \bibfield  {author} {\bibinfo {author} {\bibfnamefont {M.~J.~G.}\
  \bibnamefont {Peach}}, \bibinfo {author} {\bibfnamefont {P.}~\bibnamefont
  {Benfield}}, \bibinfo {author} {\bibfnamefont {T.}~\bibnamefont {Helgaker}},
  \ and\ \bibinfo {author} {\bibfnamefont {D.~J.}\ \bibnamefont {Tozer}},\
  }\href@noop {} {\bibfield  {journal} {\bibinfo  {journal} {J. Chem. Phys.},\
  }\textbf {\bibinfo {volume} {128}},\ \bibinfo {pages} {044118} (\bibinfo
  {year} {2008})}\BibitemShut {NoStop}%
\bibitem [{\citenamefont {Grundmann}(2010){\natexlab{a}}}]{Perovskites1}%
  \BibitemOpen
  \bibfield  {author} {\bibinfo {author} {\bibfnamefont {M.}~\bibnamefont
  {Grundmann}},\ }\href@noop {} {\emph {\bibinfo {title} {The Physics of
  Semiconductors}}}\ (\bibinfo  {publisher} {Springer-Verlag},\ \bibinfo
  {address} {Berlin},\ \bibinfo {year} {2010})\ Chap.~\bibinfo {chapter} {14},
  p.\ \bibinfo {pages} {427}\BibitemShut {NoStop}%
\bibitem [{\citenamefont {Iles}\ \emph {et~al.}(2010)\citenamefont {Iles},
  \citenamefont {Finocchi},\ and\ \citenamefont {Khodja}}]{Iles2010}%
  \BibitemOpen
  \bibfield  {author} {\bibinfo {author} {\bibfnamefont {N.}~\bibnamefont
  {Iles}}, \bibinfo {author} {\bibfnamefont {F.}~\bibnamefont {Finocchi}}, \
  and\ \bibinfo {author} {\bibfnamefont {K.~D.}\ \bibnamefont {Khodja}},\
  }\href@noop {} {\bibfield  {journal} {\bibinfo  {journal} {Journal of
  Physics: Condensed Matter},\ }\textbf {\bibinfo {volume} {22}},\ \bibinfo
  {pages} {305001} (\bibinfo {year} {2010})}\BibitemShut {NoStop}%
\bibitem [{Cha(2006)}]{Chalcopyrites2}%
  \BibitemOpen
  \href@noop {} {\emph {\bibinfo {title} {Wide-Gap Chalcopyrites}}}\ (\bibinfo
  {publisher} {Springer},\ \bibinfo {address} {Berlin},\ \bibinfo {year}
  {2006})\BibitemShut {NoStop}%
\bibitem [{Cha(1975)}]{Chalcopyrites1}%
  \BibitemOpen
  \href@noop {} {\emph {\bibinfo {title} {Ternary Chalcopyrites}}}\ (\bibinfo
  {publisher} {Pergamon},\ \bibinfo {address} {Oxford},\ \bibinfo {year}
  {1975})\BibitemShut {NoStop}%
\bibitem [{\citenamefont {Grundmann}(2010){\natexlab{b}}}]{Chalcopyrites3}%
  \BibitemOpen
  \bibfield  {author} {\bibinfo {author} {\bibfnamefont {M.}~\bibnamefont
  {Grundmann}},\ }\href@noop {} {\emph {\bibinfo {title} {The Physics of
  Semiconductors}}}\ (\bibinfo  {publisher} {Springer-Verlag},\ \bibinfo
  {address} {Berlin},\ \bibinfo {year} {2010})\ Chap.~\bibinfo {chapter} {3},
  pp.\ \bibinfo {pages} {51--53}\BibitemShut {NoStop}%
\bibitem [{\citenamefont {Gordon}(2000)}]{Delafossites2}%
  \BibitemOpen
  \bibfield  {author} {\bibinfo {author} {\bibfnamefont {R.~G.}\ \bibnamefont
  {Gordon}},\ }\href@noop {} {\bibfield  {journal} {\bibinfo  {journal} {MRS
  Bull.},\ }\textbf {\bibinfo {volume} {8}},\ \bibinfo {pages} {52} (\bibinfo
  {year} {2000})}\BibitemShut {NoStop}%
\bibitem [{\citenamefont {Grundmann}(2010){\natexlab{c}}}]{Delafossites1}%
  \BibitemOpen
  \bibfield  {author} {\bibinfo {author} {\bibfnamefont {M.}~\bibnamefont
  {Grundmann}},\ }\href@noop {} {\emph {\bibinfo {title} {The Physics of
  Semiconductors}}}\ (\bibinfo  {publisher} {Springer-Verlag},\ \bibinfo
  {address} {Berlin},\ \bibinfo {year} {2010})\ Chap.~\bibinfo {chapter} {19},
  pp.\ \bibinfo {pages} {511--515}\BibitemShut {NoStop}%
\bibitem [{\citenamefont {Perdew}\ and\ \citenamefont
  {Schmidt}(2001){\natexlab{a}}}]{JacobsLadder}%
  \BibitemOpen
  \bibfield  {author} {\bibinfo {author} {\bibfnamefont {J.~P.}\ \bibnamefont
  {Perdew}}\ and\ \bibinfo {author} {\bibfnamefont {K.}~\bibnamefont
  {Schmidt}},\ }in\ \href@noop {} {\emph {\bibinfo {booktitle} {Density
  Functional Theory and Its Applications to Materials}}},\ \bibinfo {editor}
  {edited by\ \bibinfo {editor} {\bibfnamefont {V.~V.}\ \bibnamefont
  {\textit{et al.}}}}\ (\bibinfo  {publisher} {American Institute of Physics},\
  \bibinfo {address} {New York},\ \bibinfo {year} {2001})\BibitemShut {NoStop}%
\bibitem [{\citenamefont {Perdew}\ and\ \citenamefont {Wang}(1992)}]{LSDA}%
  \BibitemOpen
  \bibfield  {author} {\bibinfo {author} {\bibfnamefont {J.~P.}\ \bibnamefont
  {Perdew}}\ and\ \bibinfo {author} {\bibfnamefont {Y.}~\bibnamefont {Wang}},\
  }\href@noop {} {\bibfield  {journal} {\bibinfo  {journal} {Phys. Rev. B},\
  }\textbf {\bibinfo {volume} {45}},\ \bibinfo {pages} {13244} (\bibinfo {year}
  {1992})}\BibitemShut {NoStop}%
\bibitem [{\citenamefont {Perdew}\ and\ \citenamefont
  {Schmidt}(2001){\natexlab{b}}}]{Perdew2001}%
  \BibitemOpen
  \bibfield  {author} {\bibinfo {author} {\bibfnamefont {J.~P.}\ \bibnamefont
  {Perdew}}\ and\ \bibinfo {author} {\bibfnamefont {K.}~\bibnamefont
  {Schmidt}}\ }(\bibinfo  {publisher} {American Institute of Physics},\
  \bibinfo {year} {2001})\BibitemShut {NoStop}%
\bibitem [{\citenamefont {Khein}\ \emph {et~al.}(1995)\citenamefont {Khein},
  \citenamefont {Singh},\ and\ \citenamefont {Umrigar}}]{Khein1995}%
  \BibitemOpen
  \bibfield  {author} {\bibinfo {author} {\bibfnamefont {A.}~\bibnamefont
  {Khein}}, \bibinfo {author} {\bibfnamefont {D.~J.}\ \bibnamefont {Singh}}, \
  and\ \bibinfo {author} {\bibfnamefont {C.}~\bibnamefont {Umrigar}},\
  }\href@noop {} {\bibfield  {journal} {\bibinfo  {journal} {Phys. Rev. B},\
  }\textbf {\bibinfo {volume} {51}},\ \bibinfo {pages} {4105} (\bibinfo {year}
  {1995})}\BibitemShut {NoStop}%
\bibitem [{\citenamefont {Kurth}\ \emph {et~al.}(1999)\citenamefont {Kurth},
  \citenamefont {Perdew},\ and\ \citenamefont {Blaha}}]{Kurth1999}%
  \BibitemOpen
  \bibfield  {author} {\bibinfo {author} {\bibfnamefont {S.}~\bibnamefont
  {Kurth}}, \bibinfo {author} {\bibfnamefont {J.~P.}\ \bibnamefont {Perdew}}, \
  and\ \bibinfo {author} {\bibfnamefont {P.}~\bibnamefont {Blaha}},\
  }\href@noop {} {\bibfield  {journal} {\bibinfo  {journal} {Int. J. Quantum
  Chem.},\ }\textbf {\bibinfo {volume} {75}},\ \bibinfo {pages} {889} (\bibinfo
  {year} {1999})}\BibitemShut {NoStop}%
\bibitem [{\citenamefont {Perdew}(1986){\natexlab{a}}}]{Perdew1986}%
  \BibitemOpen
  \bibfield  {author} {\bibinfo {author} {\bibfnamefont {J.~P.}\ \bibnamefont
  {Perdew}},\ }\href@noop {} {\bibfield  {journal} {\bibinfo  {journal} {Int.
  J. Quantum Chem. Symp},\ }\textbf {\bibinfo {volume} {19}},\ \bibinfo {pages}
  {497} (\bibinfo {year} {1986}{\natexlab{a}})}\BibitemShut {NoStop}%
\bibitem [{\citenamefont {Perdew}\ \emph {et~al.}(1996)\citenamefont {Perdew},
  \citenamefont {Burke},\ and\ \citenamefont {Ernzerhof}}]{PBE}%
  \BibitemOpen
  \bibfield  {author} {\bibinfo {author} {\bibfnamefont {J.~P.}\ \bibnamefont
  {Perdew}}, \bibinfo {author} {\bibfnamefont {K.}~\bibnamefont {Burke}}, \
  and\ \bibinfo {author} {\bibfnamefont {M.}~\bibnamefont {Ernzerhof}},\
  }\href@noop {} {\bibfield  {journal} {\bibinfo  {journal} {Phys. Rev.
  Lett.},\ }\textbf {\bibinfo {volume} {77}},\ \bibinfo {pages} {3865}
  (\bibinfo {year} {1996})},\ \bibinfo {note} {{E}\textbf{78}, 1396 (1997),
  Ref.\onlinecite{Perdew1997}}\BibitemShut {NoStop}%
\bibitem [{\citenamefont {Perdew}\ \emph {et~al.}(1997)\citenamefont {Perdew},
  \citenamefont {Burke},\ and\ \citenamefont {Ernzerhof}}]{Perdew1997}%
  \BibitemOpen
  \bibfield  {author} {\bibinfo {author} {\bibfnamefont {J.~P.}\ \bibnamefont
  {Perdew}}, \bibinfo {author} {\bibfnamefont {K.}~\bibnamefont {Burke}}, \
  and\ \bibinfo {author} {\bibfnamefont {M.}~\bibnamefont {Ernzerhof}},\
  }\href@noop {} {\bibfield  {journal} {\bibinfo  {journal} {Phys. Rev.
  Lett.},\ }\textbf {\bibinfo {volume} {78}},\ \bibinfo {pages} {1396}
  (\bibinfo {year} {1997})}\BibitemShut {NoStop}%
\bibitem [{\citenamefont {Tao}\ \emph {et~al.}(2003)\citenamefont {Tao},
  \citenamefont {Perdew}, \citenamefont {Staroverov},\ and\ \citenamefont
  {Scuseria}}]{TPSS}%
  \BibitemOpen
  \bibfield  {author} {\bibinfo {author} {\bibfnamefont {J.}~\bibnamefont
  {Tao}}, \bibinfo {author} {\bibfnamefont {J.~P.}\ \bibnamefont {Perdew}},
  \bibinfo {author} {\bibfnamefont {V.~N.}\ \bibnamefont {Staroverov}}, \ and\
  \bibinfo {author} {\bibfnamefont {G.~E.}\ \bibnamefont {Scuseria}},\
  }\href@noop {} {\bibfield  {journal} {\bibinfo  {journal} {Phys. Rev.
  Lett.},\ }\textbf {\bibinfo {volume} {91}},\ \bibinfo {pages} {146401}
  (\bibinfo {year} {2003})}\BibitemShut {NoStop}%
\bibitem [{\citenamefont {Ma}\ and\ \citenamefont {Brueckner}(1968)}]{Ma1968}%
  \BibitemOpen
  \bibfield  {author} {\bibinfo {author} {\bibfnamefont {S.-K.}\ \bibnamefont
  {Ma}}\ and\ \bibinfo {author} {\bibfnamefont {K.~A.}\ \bibnamefont
  {Brueckner}},\ }\href@noop {} {\bibfield  {journal} {\bibinfo  {journal}
  {Phys. Rev.},\ }\textbf {\bibinfo {volume} {165}},\ \bibinfo {pages} {18}
  (\bibinfo {year} {1968})}\BibitemShut {NoStop}%
\bibitem [{\citenamefont {Perdew}\ and\ \citenamefont
  {Wang}(1986)}]{Perdew1986b}%
  \BibitemOpen
  \bibfield  {author} {\bibinfo {author} {\bibfnamefont {J.~P.}\ \bibnamefont
  {Perdew}}\ and\ \bibinfo {author} {\bibfnamefont {Y.}~\bibnamefont {Wang}},\
  }\href@noop {} {\bibfield  {journal} {\bibinfo  {journal} {Phys. Rev. B},\
  }\textbf {\bibinfo {volume} {33}},\ \bibinfo {pages} {8800} (\bibinfo {year}
  {1986})},\ \bibinfo {note} {\textbf{40}, 3399(E) (1986)}\BibitemShut
  {NoStop}%
\bibitem [{\citenamefont {Perdew}(1986){\natexlab{b}}}]{Perdew1986c}%
  \BibitemOpen
  \bibfield  {author} {\bibinfo {author} {\bibfnamefont {J.~P.}\ \bibnamefont
  {Perdew}},\ }\href@noop {} {\bibfield  {journal} {\bibinfo  {journal} {Phys.
  Rev .B},\ }\textbf {\bibinfo {volume} {33}},\ \bibinfo {pages} {8822}
  (\bibinfo {year} {1986}{\natexlab{b}})},\ \bibinfo {note} {\textbf{34},
  7406(E) (1986)}\BibitemShut {NoStop}%
\bibitem [{\citenamefont {Becke}(1993){\natexlab{a}}}]{Becke1993a}%
  \BibitemOpen
  \bibfield  {author} {\bibinfo {author} {\bibfnamefont {A.~D.}\ \bibnamefont
  {Becke}},\ }\href@noop {} {\bibfield  {journal} {\bibinfo  {journal} {J.
  Chem. Phys.},\ }\textbf {\bibinfo {volume} {98}},\ \bibinfo {pages} {1372}
  (\bibinfo {year} {1993}{\natexlab{a}})}\BibitemShut {NoStop}%
\bibitem [{\citenamefont {Becke}(1993){\natexlab{b}}}]{Becke1993b}%
  \BibitemOpen
  \bibfield  {author} {\bibinfo {author} {\bibfnamefont {A.~D.}\ \bibnamefont
  {Becke}},\ }\href@noop {} {\bibfield  {journal} {\bibinfo  {journal} {J.
  Chem. Phys.},\ }\textbf {\bibinfo {volume} {98}},\ \bibinfo {pages} {5648}
  (\bibinfo {year} {1993}{\natexlab{b}})}\BibitemShut {NoStop}%
\bibitem [{\citenamefont {Stephens}\ \emph {et~al.}(1994)\citenamefont
  {Stephens}, \citenamefont {Devlin}, \citenamefont {Chabalowski},\ and\
  \citenamefont {Frisch}}]{B3LYP}%
  \BibitemOpen
  \bibfield  {author} {\bibinfo {author} {\bibfnamefont {P.~J.}\ \bibnamefont
  {Stephens}}, \bibinfo {author} {\bibfnamefont {F.~J.}\ \bibnamefont
  {Devlin}}, \bibinfo {author} {\bibfnamefont {C.~F.}\ \bibnamefont
  {Chabalowski}}, \ and\ \bibinfo {author} {\bibfnamefont {M.~J.}\ \bibnamefont
  {Frisch}},\ }\href@noop {} {\bibfield  {journal} {\bibinfo  {journal} {J.
  Phys. Chem.},\ \bibinfo {pages} {11623}} (\bibinfo {year}
  {1994})}\BibitemShut {NoStop}%
\bibitem [{\citenamefont {Adamo}\ and\ \citenamefont {Barone}(1999)}]{PBE0}%
  \BibitemOpen
  \bibfield  {author} {\bibinfo {author} {\bibfnamefont {C.}~\bibnamefont
  {Adamo}}\ and\ \bibinfo {author} {\bibfnamefont {V.}~\bibnamefont {Barone}},\
  }\href@noop {} {\bibfield  {journal} {\bibinfo  {journal} {J. Chem. Phys.},\
  }\textbf {\bibinfo {volume} {110}},\ \bibinfo {pages} {6158} (\bibinfo {year}
  {1999})}\BibitemShut {NoStop}%
\bibitem [{\citenamefont {Ernzerhof}\ and\ \citenamefont
  {Scuseria}(1999)}]{PBE1PBE}%
  \BibitemOpen
  \bibfield  {author} {\bibinfo {author} {\bibfnamefont {M.}~\bibnamefont
  {Ernzerhof}}\ and\ \bibinfo {author} {\bibfnamefont {G.~E.}\ \bibnamefont
  {Scuseria}},\ }\href@noop {} {\bibfield  {journal} {\bibinfo  {journal} {J.
  Chem. Phys.},\ }\textbf {\bibinfo {volume} {110}},\ \bibinfo {pages} {5029}
  (\bibinfo {year} {1999})}\BibitemShut {NoStop}%
\bibitem [{\citenamefont {Staroverov}\ \emph {et~al.}(2003)\citenamefont
  {Staroverov}, \citenamefont {Scuseria}, \citenamefont {Tao},\ and\
  \citenamefont {Perdew}}]{TPSSh}%
  \BibitemOpen
  \bibfield  {author} {\bibinfo {author} {\bibfnamefont {V.~N.}\ \bibnamefont
  {Staroverov}}, \bibinfo {author} {\bibfnamefont {G.~E.}\ \bibnamefont
  {Scuseria}}, \bibinfo {author} {\bibfnamefont {J.}~\bibnamefont {Tao}}, \
  and\ \bibinfo {author} {\bibfnamefont {J.~P.}\ \bibnamefont {Perdew}},\
  }\href@noop {} {\bibfield  {journal} {\bibinfo  {journal} {J. Chem. Phys.},\
  }\textbf {\bibinfo {volume} {119}},\ \bibinfo {pages} {12129} (\bibinfo
  {year} {2003})}\BibitemShut {NoStop}%
\bibitem [{\citenamefont {Zhao}\ and\ \citenamefont {Truhlar}(2008)}]{M062X}%
  \BibitemOpen
  \bibfield  {author} {\bibinfo {author} {\bibfnamefont {Y.}~\bibnamefont
  {Zhao}}\ and\ \bibinfo {author} {\bibfnamefont {D.~G.}\ \bibnamefont
  {Truhlar}},\ }\href@noop {} {\bibfield  {journal} {\bibinfo  {journal}
  {Theor. Chem. Acc.},\ }\textbf {\bibinfo {volume} {120}},\ \bibinfo {pages}
  {215} (\bibinfo {year} {2008})}\BibitemShut {NoStop}%
\bibitem [{\citenamefont {Seidl}\ \emph {et~al.}(1996)\citenamefont {Seidl},
  \citenamefont {Gorling}, \citenamefont {Vogl}, \citenamefont {Majewski},\
  and\ \citenamefont {Levy}}]{GKS}%
  \BibitemOpen
  \bibfield  {author} {\bibinfo {author} {\bibfnamefont {A.}~\bibnamefont
  {Seidl}}, \bibinfo {author} {\bibfnamefont {A.}~\bibnamefont {Gorling}},
  \bibinfo {author} {\bibfnamefont {P.}~\bibnamefont {Vogl}}, \bibinfo {author}
  {\bibfnamefont {J.~A.}\ \bibnamefont {Majewski}}, \ and\ \bibinfo {author}
  {\bibfnamefont {M.}~\bibnamefont {Levy}},\ }\href@noop {} {\bibfield
  {journal} {\bibinfo  {journal} {Phys. Rev. B},\ }\textbf {\bibinfo {volume}
  {53}},\ \bibinfo {pages} {3764} (\bibinfo {year} {1996})}\BibitemShut
  {NoStop}%
\bibitem [{\citenamefont {Henderson}\ \emph
  {et~al.}(2008){\natexlab{a}}\citenamefont {Henderson}, \citenamefont
  {Janesko},\ and\ \citenamefont {Scuseria}}]{RangeSep}%
  \BibitemOpen
  \bibfield  {author} {\bibinfo {author} {\bibfnamefont {T.~M.}\ \bibnamefont
  {Henderson}}, \bibinfo {author} {\bibfnamefont {B.~G.}\ \bibnamefont
  {Janesko}}, \ and\ \bibinfo {author} {\bibfnamefont {G.~E.}\ \bibnamefont
  {Scuseria}},\ }\href@noop {} {\bibfield  {journal} {\bibinfo  {journal} {J.
  Phys. Chem. A},\ }\textbf {\bibinfo {volume} {112}},\ \bibinfo {pages}
  {12530} (\bibinfo {year} {2008}{\natexlab{a}})}\BibitemShut {NoStop}%
\bibitem [{\citenamefont {Paier}\ \emph {et~al.}(2006)\citenamefont {Paier},
  \citenamefont {Marsman}, \citenamefont {Kresse}, \citenamefont {Gerber},\
  and\ \citenamefont {\'Angy\'an}}]{Paier2006}%
  \BibitemOpen
  \bibfield  {author} {\bibinfo {author} {\bibfnamefont {J.}~\bibnamefont
  {Paier}}, \bibinfo {author} {\bibfnamefont {M.}~\bibnamefont {Marsman}},
  \bibinfo {author} {\bibfnamefont {G.}~\bibnamefont {Kresse}}, \bibinfo
  {author} {\bibfnamefont {I.~C.}\ \bibnamefont {Gerber}}, \ and\ \bibinfo
  {author} {\bibfnamefont {J.~G.}\ \bibnamefont {\'Angy\'an}},\ }\href@noop {}
  {\bibfield  {journal} {\bibinfo  {journal} {J. Chem. Phys.},\ }\textbf
  {\bibinfo {volume} {124}},\ \bibinfo {pages} {154709} (\bibinfo {year}
  {2006})}\BibitemShut {NoStop}%
\bibitem [{\citenamefont {Savin}(1996)}]{SavinBook}%
  \BibitemOpen
  \bibfield  {author} {\bibinfo {author} {\bibfnamefont {A.}~\bibnamefont
  {Savin}},\ }in\ \href@noop {} {\emph {\bibinfo {booktitle} {Recent
  Developments and Applications of Modern Density Functional Theory}}},\
  \bibinfo {editor} {edited by\ \bibinfo {editor} {\bibfnamefont {J.~M.}\
  \bibnamefont {Seminario}}}\ (\bibinfo  {publisher} {Elsevier},\ \bibinfo
  {address} {Amsterdam, The Netherlands},\ \bibinfo {year} {1996})\ pp.\
  \bibinfo {pages} {327--357}\BibitemShut {NoStop}%
\bibitem [{\citenamefont {Savin}\ and\ \citenamefont {Flad}(1995)}]{Savin1995}%
  \BibitemOpen
  \bibfield  {author} {\bibinfo {author} {\bibfnamefont {A.}~\bibnamefont
  {Savin}}\ and\ \bibinfo {author} {\bibfnamefont {H.-J.}\ \bibnamefont
  {Flad}},\ }\href@noop {} {\bibfield  {journal} {\bibinfo  {journal} {Int. J.
  Quantum Chem.},\ }\textbf {\bibinfo {volume} {56}},\ \bibinfo {pages} {327}
  (\bibinfo {year} {1995})}\BibitemShut {NoStop}%
\bibitem [{\citenamefont {Leininger}\ \emph {et~al.}(1997)\citenamefont
  {Leininger}, \citenamefont {Stoll}, \citenamefont {Werner},\ and\
  \citenamefont {Savin}}]{Savin1997}%
  \BibitemOpen
  \bibfield  {author} {\bibinfo {author} {\bibfnamefont {T.}~\bibnamefont
  {Leininger}}, \bibinfo {author} {\bibfnamefont {H.}~\bibnamefont {Stoll}},
  \bibinfo {author} {\bibfnamefont {H.-J.}\ \bibnamefont {Werner}}, \ and\
  \bibinfo {author} {\bibfnamefont {A.}~\bibnamefont {Savin}},\ }\href@noop {}
  {\bibfield  {journal} {\bibinfo  {journal} {Chem. Phys. Lett.},\ }\textbf
  {\bibinfo {volume} {275}},\ \bibinfo {pages} {151} (\bibinfo {year}
  {1997})}\BibitemShut {NoStop}%
\bibitem [{\citenamefont {Vydrov}\ and\ \citenamefont
  {Scuseria}(2006)}]{LCwPBE}%
  \BibitemOpen
  \bibfield  {author} {\bibinfo {author} {\bibfnamefont {O.~A.}\ \bibnamefont
  {Vydrov}}\ and\ \bibinfo {author} {\bibfnamefont {G.~E.}\ \bibnamefont
  {Scuseria}},\ }\href@noop {} {\bibfield  {journal} {\bibinfo  {journal} {J.
  Chem. Phys.},\ }\textbf {\bibinfo {volume} {125}},\ \bibinfo {pages} {234109}
  (\bibinfo {year} {2006})}\BibitemShut {NoStop}%
\bibitem [{\citenamefont {Iikura}\ \emph {et~al.}(2001)\citenamefont {Iikura},
  \citenamefont {Tsuneda}, \citenamefont {Yanai},\ and\ \citenamefont
  {Hirao}}]{Hirao2001}%
  \BibitemOpen
  \bibfield  {author} {\bibinfo {author} {\bibfnamefont {H.}~\bibnamefont
  {Iikura}}, \bibinfo {author} {\bibfnamefont {T.}~\bibnamefont {Tsuneda}},
  \bibinfo {author} {\bibfnamefont {T.}~\bibnamefont {Yanai}}, \ and\ \bibinfo
  {author} {\bibfnamefont {K.}~\bibnamefont {Hirao}},\ }\href@noop {}
  {\bibfield  {journal} {\bibinfo  {journal} {J. Chem. Phys.},\ }\textbf
  {\bibinfo {volume} {115}},\ \bibinfo {pages} {3540} (\bibinfo {year}
  {2001})}\BibitemShut {NoStop}%
\bibitem [{\citenamefont {Sekino}\ \emph {et~al.}(2005)\citenamefont {Sekino},
  \citenamefont {Maeda},\ and\ \citenamefont {Kamiya}}]{Sekino2005}%
  \BibitemOpen
  \bibfield  {author} {\bibinfo {author} {\bibfnamefont {H.}~\bibnamefont
  {Sekino}}, \bibinfo {author} {\bibfnamefont {Y.}~\bibnamefont {Maeda}}, \
  and\ \bibinfo {author} {\bibfnamefont {M.}~\bibnamefont {Kamiya}},\
  }\href@noop {} {\bibfield  {journal} {\bibinfo  {journal} {Mol. Phys.},\
  }\textbf {\bibinfo {volume} {103}},\ \bibinfo {pages} {2183} (\bibinfo {year}
  {2005})}\BibitemShut {NoStop}%
\bibitem [{\citenamefont {Heyd}\ \emph {et~al.}(2003)\citenamefont {Heyd},
  \citenamefont {Scuseria},\ and\ \citenamefont {Ernzerhof}}]{HSE03}%
  \BibitemOpen
  \bibfield  {author} {\bibinfo {author} {\bibfnamefont {J.}~\bibnamefont
  {Heyd}}, \bibinfo {author} {\bibfnamefont {G.~E.}\ \bibnamefont {Scuseria}},
  \ and\ \bibinfo {author} {\bibfnamefont {M.}~\bibnamefont {Ernzerhof}},\
  }\href@noop {} {\bibfield  {journal} {\bibinfo  {journal} {J. Chem. Phys.},\
  }\textbf {\bibinfo {volume} {118}},\ \bibinfo {pages} {8207} (\bibinfo {year}
  {2003})}\BibitemShut {NoStop}%
\bibitem [{\citenamefont {Krukau}\ \emph {et~al.}(2006)\citenamefont {Krukau},
  \citenamefont {Vydrov}, \citenamefont {Izmaylov},\ and\ \citenamefont
  {Scuseria}}]{HSE06}%
  \BibitemOpen
  \bibfield  {author} {\bibinfo {author} {\bibfnamefont {A.~V.}\ \bibnamefont
  {Krukau}}, \bibinfo {author} {\bibfnamefont {O.~A.}\ \bibnamefont {Vydrov}},
  \bibinfo {author} {\bibfnamefont {A.~F.}\ \bibnamefont {Izmaylov}}, \ and\
  \bibinfo {author} {\bibfnamefont {G.~E.}\ \bibnamefont {Scuseria}},\
  }\href@noop {} {\bibfield  {journal} {\bibinfo  {journal} {J. Chem. Phys.},\
  }\textbf {\bibinfo {volume} {125}},\ \bibinfo {pages} {224106} (\bibinfo
  {year} {2006})}\BibitemShut {NoStop}%
\bibitem [{\citenamefont {Henderson}\ \emph {et~al.}(2009)\citenamefont
  {Henderson}, \citenamefont {Izmaylov}, \citenamefont {Scalmani},\ and\
  \citenamefont {Scuseria}}]{HSEh}%
  \BibitemOpen
  \bibfield  {author} {\bibinfo {author} {\bibfnamefont {T.~M.}\ \bibnamefont
  {Henderson}}, \bibinfo {author} {\bibfnamefont {A.~F.}\ \bibnamefont
  {Izmaylov}}, \bibinfo {author} {\bibfnamefont {G.}~\bibnamefont {Scalmani}},
  \ and\ \bibinfo {author} {\bibfnamefont {G.~E.}\ \bibnamefont {Scuseria}},\
  }\href@noop {} {\bibfield  {journal} {\bibinfo  {journal} {J. Chem. Phys.},\
  }\textbf {\bibinfo {volume} {131}},\ \bibinfo {pages} {044108} (\bibinfo
  {year} {2009})},\ \bibinfo {note} {specifically, the {HSEh} parameterization
  of {HSE06}\cite{HSE06,HSE06e}, called by the {\sc{gaussian}} keyword
  {HSEh1PBE}.}\BibitemShut {Stop}%
\bibitem [{\citenamefont {Janesko}\ \emph {et~al.}(2009)\citenamefont
  {Janesko}, \citenamefont {Henderson},\ and\ \citenamefont
  {Scuseria}}]{Janesko2009}%
  \BibitemOpen
  \bibfield  {author} {\bibinfo {author} {\bibfnamefont {B.~G.}\ \bibnamefont
  {Janesko}}, \bibinfo {author} {\bibfnamefont {T.~M.}\ \bibnamefont
  {Henderson}}, \ and\ \bibinfo {author} {\bibfnamefont {G.~E.}\ \bibnamefont
  {Scuseria}},\ }\href@noop {} {\bibfield  {journal} {\bibinfo  {journal}
  {Phys. Chem. Chem. Phys.},\ }\textbf {\bibinfo {volume} {11}},\ \bibinfo
  {pages} {443} (\bibinfo {year} {2009})}\BibitemShut {NoStop}%
\bibitem [{\citenamefont {Barone}\ \emph {et~al.}(2011)\citenamefont {Barone},
  \citenamefont {Hod}, \citenamefont {Peralta},\ and\ \citenamefont
  {Scuseria}}]{Barone2011}%
  \BibitemOpen
  \bibfield  {author} {\bibinfo {author} {\bibfnamefont {V.}~\bibnamefont
  {Barone}}, \bibinfo {author} {\bibfnamefont {O.}~\bibnamefont {Hod}},
  \bibinfo {author} {\bibfnamefont {J.~E.}\ \bibnamefont {Peralta}}, \ and\
  \bibinfo {author} {\bibfnamefont {G.~E.}\ \bibnamefont {Scuseria}},\
  }\href@noop {} {\bibfield  {journal} {\bibinfo  {journal} {Acc. Chem. Res.},\
  }\textbf {\bibinfo {volume} {44}},\ \bibinfo {pages} {269} (\bibinfo {year}
  {2011})}\BibitemShut {NoStop}%
\bibitem [{\citenamefont {Henderson}\ \emph {et~al.}(2007)\citenamefont
  {Henderson}, \citenamefont {Izmaylov}, \citenamefont {Scuseria},\ and\
  \citenamefont {Savin}}]{HISS1}%
  \BibitemOpen
  \bibfield  {author} {\bibinfo {author} {\bibfnamefont {T.~M.}\ \bibnamefont
  {Henderson}}, \bibinfo {author} {\bibfnamefont {A.~F.}\ \bibnamefont
  {Izmaylov}}, \bibinfo {author} {\bibfnamefont {G.~E.}\ \bibnamefont
  {Scuseria}}, \ and\ \bibinfo {author} {\bibfnamefont {A.}~\bibnamefont
  {Savin}},\ }\href@noop {} {\bibfield  {journal} {\bibinfo  {journal} {J.
  Chem. Phys.},\ }\textbf {\bibinfo {volume} {127}},\ \bibinfo {pages} {221103}
  (\bibinfo {year} {2007})}\BibitemShut {NoStop}%
\bibitem [{\citenamefont {Henderson}\ \emph
  {et~al.}(2008){\natexlab{b}}\citenamefont {Henderson}, \citenamefont
  {Izmaylov}, \citenamefont {Scuseria},\ and\ \citenamefont {Savin}}]{HISS2}%
  \BibitemOpen
  \bibfield  {author} {\bibinfo {author} {\bibfnamefont {T.~M.}\ \bibnamefont
  {Henderson}}, \bibinfo {author} {\bibfnamefont {A.~F.}\ \bibnamefont
  {Izmaylov}}, \bibinfo {author} {\bibfnamefont {G.~E.}\ \bibnamefont
  {Scuseria}}, \ and\ \bibinfo {author} {\bibfnamefont {A.}~\bibnamefont
  {Savin}},\ }\href@noop {} {\bibfield  {journal} {\bibinfo  {journal} {J.
  Theor. Comput. Chem.},\ }\textbf {\bibinfo {volume} {4}},\ \bibinfo {pages}
  {1254} (\bibinfo {year} {2008}{\natexlab{b}})}\BibitemShut {NoStop}%
\bibitem [{\citenamefont {Kudin}\ and\ \citenamefont
  {Scuseria}(2000)}]{Kudin2000}%
  \BibitemOpen
  \bibfield  {author} {\bibinfo {author} {\bibfnamefont {K.~N.}\ \bibnamefont
  {Kudin}}\ and\ \bibinfo {author} {\bibfnamefont {G.~E.}\ \bibnamefont
  {Scuseria}},\ }\href@noop {} {\bibfield  {journal} {\bibinfo  {journal}
  {Phys. Rev. B},\ }\textbf {\bibinfo {volume} {61}},\ \bibinfo {pages} {16440}
  (\bibinfo {year} {2000})}\BibitemShut {NoStop}%
\bibitem [{\citenamefont {Kudin}\ and\ \citenamefont
  {Scuseria}(1998)}]{Kudin1998b}%
  \BibitemOpen
  \bibfield  {author} {\bibinfo {author} {\bibfnamefont {K.~N.}\ \bibnamefont
  {Kudin}}\ and\ \bibinfo {author} {\bibfnamefont {G.~E.}\ \bibnamefont
  {Scuseria}},\ }\href@noop {} {\bibfield  {journal} {\bibinfo  {journal}
  {Chem. Phys. Lett.},\ }\textbf {\bibinfo {volume} {289}},\ \bibinfo {pages}
  {611} (\bibinfo {year} {1998})}\BibitemShut {NoStop}%
\bibitem [{\citenamefont {Kudin}\ and\ \citenamefont
  {E.Scuseria}(1998)}]{Kudin1998a}%
  \BibitemOpen
  \bibfield  {author} {\bibinfo {author} {\bibfnamefont {K.~N.}\ \bibnamefont
  {Kudin}}\ and\ \bibinfo {author} {\bibfnamefont {G.}~\bibnamefont
  {E.Scuseria}},\ }\href@noop {} {\bibfield  {journal} {\bibinfo  {journal}
  {Chem. Phys. Lett.},\ }\textbf {\bibinfo {volume} {283}},\ \bibinfo {pages}
  {61} (\bibinfo {year} {1998})}\BibitemShut {NoStop}%
\bibitem [{\citenamefont {Frisch}\ \emph {et~al.}(2009)\citenamefont {Frisch},
  \citenamefont {Trucks}, \citenamefont {Schlegel}, \citenamefont {Scuseria},
  \citenamefont {Robb}, \citenamefont {Cheeseman}, \citenamefont {Scalmani},
  \citenamefont {Barone}, \citenamefont {Mennucci}, \citenamefont {Petersson},
  \citenamefont {Nakatsuji}, \citenamefont {Caricato}, \citenamefont {Li},
  \citenamefont {Hratchian}, \citenamefont {Izmaylov}, \citenamefont {Bloino},
  \citenamefont {Zheng}, \citenamefont {Sonnenberg}, \citenamefont {Hada},
  \citenamefont {Ehara}, \citenamefont {Toyota}, \citenamefont {Fukuda},
  \citenamefont {Hasegawa}, \citenamefont {Ishida}, \citenamefont {Nakajima},
  \citenamefont {Honda}, \citenamefont {Kitao}, \citenamefont {Nakai},
  \citenamefont {Vreven}, \citenamefont {Montgomery}, \citenamefont {Peralta},
  \citenamefont {Ogliaro}, \citenamefont {Bearpark}, \citenamefont {Heyd},
  \citenamefont {Brothers}, \citenamefont {Kudin}, \citenamefont {Staroverov},
  \citenamefont {Kobayashi}, \citenamefont {Normand}, \citenamefont
  {Raghavachari}, \citenamefont {Rendell}, \citenamefont {Burant},
  \citenamefont {Iyengar}, \citenamefont {Tomasi}, \citenamefont {Cossi},
  \citenamefont {Rega}, \citenamefont {Millam}, \citenamefont {Klene},
  \citenamefont {Knox}, \citenamefont {Cross}, \citenamefont {Bakken},
  \citenamefont {Adamo}, \citenamefont {Jaramillo}, \citenamefont {Gomperts},
  \citenamefont {Stratmann}, \citenamefont {Yazyev}, \citenamefont {Austin},
  \citenamefont {Cammi}, \citenamefont {Pomelli}, \citenamefont {Ochterski},
  \citenamefont {Martin}, \citenamefont {Morokuma}, \citenamefont {Zakrzewski},
  \citenamefont {Voth}, \citenamefont {Salvador}, \citenamefont {Dannenberg},
  \citenamefont {Dapprich}, \citenamefont {Daniels}, \citenamefont {Farkas},
  \citenamefont {Foresman}, \citenamefont {Ortiz}, \citenamefont {Cioslowski},\
  and\ \citenamefont {Fox}}]{G09}%
  \BibitemOpen
  \bibfield  {author} {\bibinfo {author} {\bibfnamefont {M.~J.}\ \bibnamefont
  {Frisch}}, \bibinfo {author} {\bibfnamefont {G.~W.}\ \bibnamefont {Trucks}},
  \bibinfo {author} {\bibfnamefont {H.~B.}\ \bibnamefont {Schlegel}}, \bibinfo
  {author} {\bibfnamefont {G.~E.}\ \bibnamefont {Scuseria}}, \bibinfo {author}
  {\bibfnamefont {M.~A.}\ \bibnamefont {Robb}}, \bibinfo {author}
  {\bibfnamefont {J.~R.}\ \bibnamefont {Cheeseman}}, \bibinfo {author}
  {\bibfnamefont {G.}~\bibnamefont {Scalmani}}, \bibinfo {author}
  {\bibfnamefont {V.}~\bibnamefont {Barone}}, \bibinfo {author} {\bibfnamefont
  {B.}~\bibnamefont {Mennucci}}, \bibinfo {author} {\bibfnamefont {G.~A.}\
  \bibnamefont {Petersson}}, \bibinfo {author} {\bibfnamefont {H.}~\bibnamefont
  {Nakatsuji}}, \bibinfo {author} {\bibfnamefont {M.}~\bibnamefont {Caricato}},
  \bibinfo {author} {\bibfnamefont {X.}~\bibnamefont {Li}}, \bibinfo {author}
  {\bibfnamefont {H.~P.}\ \bibnamefont {Hratchian}}, \bibinfo {author}
  {\bibfnamefont {A.~F.}\ \bibnamefont {Izmaylov}}, \bibinfo {author}
  {\bibfnamefont {J.}~\bibnamefont {Bloino}}, \bibinfo {author} {\bibfnamefont
  {G.}~\bibnamefont {Zheng}}, \bibinfo {author} {\bibfnamefont {J.~L.}\
  \bibnamefont {Sonnenberg}}, \bibinfo {author} {\bibfnamefont
  {M.}~\bibnamefont {Hada}}, \bibinfo {author} {\bibfnamefont {M.}~\bibnamefont
  {Ehara}}, \bibinfo {author} {\bibfnamefont {K.}~\bibnamefont {Toyota}},
  \bibinfo {author} {\bibfnamefont {R.}~\bibnamefont {Fukuda}}, \bibinfo
  {author} {\bibfnamefont {J.}~\bibnamefont {Hasegawa}}, \bibinfo {author}
  {\bibfnamefont {M.}~\bibnamefont {Ishida}}, \bibinfo {author} {\bibfnamefont
  {T.}~\bibnamefont {Nakajima}}, \bibinfo {author} {\bibfnamefont
  {Y.}~\bibnamefont {Honda}}, \bibinfo {author} {\bibfnamefont
  {O.}~\bibnamefont {Kitao}}, \bibinfo {author} {\bibfnamefont
  {H.}~\bibnamefont {Nakai}}, \bibinfo {author} {\bibfnamefont
  {T.}~\bibnamefont {Vreven}}, \bibinfo {author} {\bibfnamefont {J.~A.}\
  \bibnamefont {Montgomery}, \bibfnamefont {{Jr.}}}, \bibinfo {author}
  {\bibfnamefont {J.~E.}\ \bibnamefont {Peralta}}, \bibinfo {author}
  {\bibfnamefont {F.}~\bibnamefont {Ogliaro}}, \bibinfo {author} {\bibfnamefont
  {M.}~\bibnamefont {Bearpark}}, \bibinfo {author} {\bibfnamefont {J.~J.}\
  \bibnamefont {Heyd}}, \bibinfo {author} {\bibfnamefont {E.}~\bibnamefont
  {Brothers}}, \bibinfo {author} {\bibfnamefont {K.~N.}\ \bibnamefont {Kudin}},
  \bibinfo {author} {\bibfnamefont {V.~N.}\ \bibnamefont {Staroverov}},
  \bibinfo {author} {\bibfnamefont {R.}~\bibnamefont {Kobayashi}}, \bibinfo
  {author} {\bibfnamefont {J.}~\bibnamefont {Normand}}, \bibinfo {author}
  {\bibfnamefont {K.}~\bibnamefont {Raghavachari}}, \bibinfo {author}
  {\bibfnamefont {A.}~\bibnamefont {Rendell}}, \bibinfo {author} {\bibfnamefont
  {J.~C.}\ \bibnamefont {Burant}}, \bibinfo {author} {\bibfnamefont {S.~S.}\
  \bibnamefont {Iyengar}}, \bibinfo {author} {\bibfnamefont {J.}~\bibnamefont
  {Tomasi}}, \bibinfo {author} {\bibfnamefont {M.}~\bibnamefont {Cossi}},
  \bibinfo {author} {\bibfnamefont {N.}~\bibnamefont {Rega}}, \bibinfo {author}
  {\bibfnamefont {J.~M.}\ \bibnamefont {Millam}}, \bibinfo {author}
  {\bibfnamefont {M.}~\bibnamefont {Klene}}, \bibinfo {author} {\bibfnamefont
  {J.~E.}\ \bibnamefont {Knox}}, \bibinfo {author} {\bibfnamefont {J.~B.}\
  \bibnamefont {Cross}}, \bibinfo {author} {\bibfnamefont {V.}~\bibnamefont
  {Bakken}}, \bibinfo {author} {\bibfnamefont {C.}~\bibnamefont {Adamo}},
  \bibinfo {author} {\bibfnamefont {J.}~\bibnamefont {Jaramillo}}, \bibinfo
  {author} {\bibfnamefont {R.}~\bibnamefont {Gomperts}}, \bibinfo {author}
  {\bibfnamefont {R.~E.}\ \bibnamefont {Stratmann}}, \bibinfo {author}
  {\bibfnamefont {O.}~\bibnamefont {Yazyev}}, \bibinfo {author} {\bibfnamefont
  {A.~J.}\ \bibnamefont {Austin}}, \bibinfo {author} {\bibfnamefont
  {R.}~\bibnamefont {Cammi}}, \bibinfo {author} {\bibfnamefont
  {C.}~\bibnamefont {Pomelli}}, \bibinfo {author} {\bibfnamefont {J.~W.}\
  \bibnamefont {Ochterski}}, \bibinfo {author} {\bibfnamefont {R.~L.}\
  \bibnamefont {Martin}}, \bibinfo {author} {\bibfnamefont {K.}~\bibnamefont
  {Morokuma}}, \bibinfo {author} {\bibfnamefont {V.~G.}\ \bibnamefont
  {Zakrzewski}}, \bibinfo {author} {\bibfnamefont {G.~A.}\ \bibnamefont
  {Voth}}, \bibinfo {author} {\bibfnamefont {P.}~\bibnamefont {Salvador}},
  \bibinfo {author} {\bibfnamefont {J.~J.}\ \bibnamefont {Dannenberg}},
  \bibinfo {author} {\bibfnamefont {S.}~\bibnamefont {Dapprich}}, \bibinfo
  {author} {\bibfnamefont {A.~D.}\ \bibnamefont {Daniels}}, \bibinfo {author}
  {\bibfnamefont {O.}~\bibnamefont {Farkas}}, \bibinfo {author} {\bibfnamefont
  {J.~B.}\ \bibnamefont {Foresman}}, \bibinfo {author} {\bibfnamefont {J.~V.}\
  \bibnamefont {Ortiz}}, \bibinfo {author} {\bibfnamefont {J.}~\bibnamefont
  {Cioslowski}}, \ and\ \bibinfo {author} {\bibfnamefont {D.~J.}\ \bibnamefont
  {Fox}},\ }\href@noop {} {\enquote {\bibinfo {title} {Gaussian~09 {R}evision
  {B}.1 and gdv-h11},}\ } (\bibinfo {year} {2009}),\ \bibinfo {note}
  {{G}aussian Inc., Wallingford CT}\BibitemShut {NoStop}%
\bibitem [{HSE()}]{HSEepaps}%
  \BibitemOpen
  \Doi {doi:10.1063/1.2085170} {}\bibinfo {note} {The {EPAPS} {D}ocument {N}o.
  E-JCPSA6-123-301539 contains modified basis sets, input files and geometries
  relevant to the material in Ref.~\onlinecite{Heyd2005}. {P}lease see
  \url{doi:10.1063/1.2085170} for the direct link or
  \url{http://www.aip.org/pubservs/epaps.html} for more information on
  {EPAPS}.}\BibitemShut {Stop}%
\bibitem [{\citenamefont {Heyd}\ \emph {et~al.}(2005)\citenamefont {Heyd},
  \citenamefont {Peralta}, \citenamefont {Scuseria},\ and\ \citenamefont
  {Martin}}]{Heyd2005}%
  \BibitemOpen
  \bibfield  {author} {\bibinfo {author} {\bibfnamefont {J.}~\bibnamefont
  {Heyd}}, \bibinfo {author} {\bibfnamefont {J.~E.}\ \bibnamefont {Peralta}},
  \bibinfo {author} {\bibfnamefont {G.~E.}\ \bibnamefont {Scuseria}}, \ and\
  \bibinfo {author} {\bibfnamefont {R.~L.}\ \bibnamefont {Martin}},\
  }\href@noop {} {\bibfield  {journal} {\bibinfo  {journal} {J. Chem. Phys.},\
  }\textbf {\bibinfo {volume} {123}},\ \bibinfo {pages} {174101} (\bibinfo
  {year} {2005})}\BibitemShut {NoStop}%
\bibitem [{\citenamefont {Vosko}\ \emph {et~al.}(1980)\citenamefont {Vosko},
  \citenamefont {Wilk},\ and\ \citenamefont {Nusair}}]{SVWN5}%
  \BibitemOpen
  \bibfield  {author} {\bibinfo {author} {\bibfnamefont {S.~H.}\ \bibnamefont
  {Vosko}}, \bibinfo {author} {\bibfnamefont {L.}~\bibnamefont {Wilk}}, \ and\
  \bibinfo {author} {\bibfnamefont {M.}~\bibnamefont {Nusair}},\ }\href@noop {}
  {\bibfield  {journal} {\bibinfo  {journal} {Can. J. Phys},\ }\textbf
  {\bibinfo {volume} {58}},\ \bibinfo {pages} {1200} (\bibinfo {year}
  {1980})}\BibitemShut {NoStop}%
\bibitem [{\citenamefont {Perdew}\ \emph {et~al.}(2008)\citenamefont {Perdew},
  \citenamefont {Ruzsinszky}, \citenamefont {Csonka}, \citenamefont {Vydrov},
  \citenamefont {Scuseria}, \citenamefont {Constantin}, \citenamefont {Zhou},\
  and\ \citenamefont {Burke}}]{PBEsol}%
  \BibitemOpen
  \bibfield  {author} {\bibinfo {author} {\bibfnamefont {J.~P.}\ \bibnamefont
  {Perdew}}, \bibinfo {author} {\bibfnamefont {A.}~\bibnamefont {Ruzsinszky}},
  \bibinfo {author} {\bibfnamefont {G.~I.}\ \bibnamefont {Csonka}}, \bibinfo
  {author} {\bibfnamefont {O.~A.}\ \bibnamefont {Vydrov}}, \bibinfo {author}
  {\bibfnamefont {G.~E.}\ \bibnamefont {Scuseria}}, \bibinfo {author}
  {\bibfnamefont {L.~A.}\ \bibnamefont {Constantin}}, \bibinfo {author}
  {\bibfnamefont {X.}~\bibnamefont {Zhou}}, \ and\ \bibinfo {author}
  {\bibfnamefont {K.}~\bibnamefont {Burke}},\ }\Doi
  {10.1103/PhysRevLett.100.136406} {\bibfield  {journal} {\bibinfo  {journal}
  {Phys. Rev. Lett.},\ }\textbf {\bibinfo {volume} {100}},\ \bibinfo {pages}
  {136406} (\bibinfo {year} {2008})},\ \bibinfo {note} {see also the 2009
  Erratum in Reference\cite{PBEsolErr}}\BibitemShut {NoStop}%
\bibitem [{\citenamefont {Kudin}\ \emph {et~al.}(2001)\citenamefont {Kudin},
  \citenamefont {Scuseria},\ and\ \citenamefont {Schlegel}}]{Kudin2001}%
  \BibitemOpen
  \bibfield  {author} {\bibinfo {author} {\bibfnamefont {K.~N.}\ \bibnamefont
  {Kudin}}, \bibinfo {author} {\bibfnamefont {G.~E.}\ \bibnamefont {Scuseria}},
  \ and\ \bibinfo {author} {\bibfnamefont {H.~B.}\ \bibnamefont {Schlegel}},\
  }\href@noop {} {\bibfield  {journal} {\bibinfo  {journal} {J. Chem. Phys.},\
  }\textbf {\bibinfo {volume} {114}},\ \bibinfo {pages} {2919} (\bibinfo {year}
  {2001})}\BibitemShut {NoStop}%
\bibitem [{\citenamefont {Madelung}(2004)}]{Madelung2004}%
  \BibitemOpen
  \bibfield  {author} {\bibinfo {author} {\bibfnamefont {O.}~\bibnamefont
  {Madelung}},\ }\href@noop {} {\emph {\bibinfo {title} {Semiconductors - Data
  Handbook 3rd Ed.}}}\ (\bibinfo  {publisher} {Springer},\ \bibinfo {address}
  {New York},\ \bibinfo {year} {2004})\BibitemShut {NoStop}%
\bibitem [{\citenamefont {Arnaudov}\ \emph {et~al.}(2004)\citenamefont
  {Arnaudov}, \citenamefont {Paskova}, \citenamefont {Paskov}, \citenamefont
  {Magnusson}, \citenamefont {Valcheva}, \citenamefont {Monemar}, \citenamefont
  {Lu}, \citenamefont {Schaff}, \citenamefont {Amano},\ and\ \citenamefont
  {Akasaki}}]{Arnaudov2004}%
  \BibitemOpen
  \bibfield  {author} {\bibinfo {author} {\bibfnamefont {B.}~\bibnamefont
  {Arnaudov}}, \bibinfo {author} {\bibfnamefont {T.}~\bibnamefont {Paskova}},
  \bibinfo {author} {\bibfnamefont {P.~P.}\ \bibnamefont {Paskov}}, \bibinfo
  {author} {\bibfnamefont {B.}~\bibnamefont {Magnusson}}, \bibinfo {author}
  {\bibfnamefont {E.}~\bibnamefont {Valcheva}}, \bibinfo {author}
  {\bibfnamefont {B.}~\bibnamefont {Monemar}}, \bibinfo {author} {\bibfnamefont
  {H.}~\bibnamefont {Lu}}, \bibinfo {author} {\bibfnamefont {W.~J.}\
  \bibnamefont {Schaff}}, \bibinfo {author} {\bibfnamefont {H.}~\bibnamefont
  {Amano}}, \ and\ \bibinfo {author} {\bibfnamefont {I.}~\bibnamefont
  {Akasaki}},\ }\Doi {10.1103/PhysRevB.69.115216} {\bibfield  {journal}
  {\bibinfo  {journal} {Phys. Rev. B},\ }\textbf {\bibinfo {volume} {69}},\
  \bibinfo {pages} {115216} (\bibinfo {year} {2004})}\BibitemShut {NoStop}%
\bibitem [{\citenamefont {Yacobi}(2003)}]{Yacobi2003}%
  \BibitemOpen
  \bibfield  {author} {\bibinfo {author} {\bibfnamefont {B.~G.}\ \bibnamefont
  {Yacobi}},\ }\href@noop {} {\emph {\bibinfo {title} {Semiconductor Materials:
  An Introduction to Basic Principles}}}\ (\bibinfo  {publisher} {Kluwer
  Academic/Plenum Publishers},\ \bibinfo {address} {New York},\ \bibinfo {year}
  {2003})\BibitemShut {NoStop}%
\bibitem [{\citenamefont {Doorn}\ and\ \citenamefont
  {Nobel}(1956)}]{vanDoorn1956}%
  \BibitemOpen
  \bibfield  {author} {\bibinfo {author} {\bibfnamefont {C.~V.}\ \bibnamefont
  {Doorn}}\ and\ \bibinfo {author} {\bibfnamefont {D.~D.}\ \bibnamefont
  {Nobel}},\ }\href@noop {} {\bibfield  {journal} {\bibinfo  {journal}
  {Physica},\ }\textbf {\bibinfo {volume} {22}},\ \bibinfo {pages} {338 }
  (\bibinfo {year} {1956})}\BibitemShut {NoStop}%
\bibitem [{\citenamefont {Davidson}\ \emph {et~al.}(2010)\citenamefont
  {Davidson}, \citenamefont {Moug}, \citenamefont {Izdebski}, \citenamefont
  {Bradford},\ and\ \citenamefont {Prior}}]{Davidson2010}%
  \BibitemOpen
  \bibfield  {author} {\bibinfo {author} {\bibfnamefont {I.~A.}\ \bibnamefont
  {Davidson}}, \bibinfo {author} {\bibfnamefont {R.~T.}\ \bibnamefont {Moug}},
  \bibinfo {author} {\bibfnamefont {F.}~\bibnamefont {Izdebski}}, \bibinfo
  {author} {\bibfnamefont {C.}~\bibnamefont {Bradford}}, \ and\ \bibinfo
  {author} {\bibfnamefont {K.~A.}\ \bibnamefont {Prior}},\ }\Doi
  {10.1002/pssb.200983190} {\bibfield  {journal} {\bibinfo  {journal} {physica
  status solidi (b)},\ }\textbf {\bibinfo {volume} {247}},\ \bibinfo {pages}
  {1396} (\bibinfo {year} {2010})}\BibitemShut {NoStop}%
\bibitem [{\citenamefont {Hartmann}\ \emph {et~al.}(1996)\citenamefont
  {Hartmann}, \citenamefont {Cibert}, \citenamefont {Kany}, \citenamefont
  {Mariette}, \citenamefont {Charleux}, \citenamefont {Alleysson},
  \citenamefont {Langer},\ and\ \citenamefont {Feuillet}}]{Hartmann1996}%
  \BibitemOpen
  \bibfield  {author} {\bibinfo {author} {\bibfnamefont {J.~M.}\ \bibnamefont
  {Hartmann}}, \bibinfo {author} {\bibfnamefont {J.}~\bibnamefont {Cibert}},
  \bibinfo {author} {\bibfnamefont {F.}~\bibnamefont {Kany}}, \bibinfo {author}
  {\bibfnamefont {H.}~\bibnamefont {Mariette}}, \bibinfo {author}
  {\bibfnamefont {M.}~\bibnamefont {Charleux}}, \bibinfo {author}
  {\bibfnamefont {P.}~\bibnamefont {Alleysson}}, \bibinfo {author}
  {\bibfnamefont {R.}~\bibnamefont {Langer}}, \ and\ \bibinfo {author}
  {\bibfnamefont {G.}~\bibnamefont {Feuillet}},\ }\Doi {DOI:10.1063/1.363714}
  {\textbf {\bibinfo {volume} {80}},\ \bibinfo {pages} {6257} (\bibinfo {year}
  {1996})}\BibitemShut {NoStop}%
\bibitem [{And(2006)}]{Anderson2006}%
  \BibitemOpen
  \Doi {doi:10.1088/0964-1726/15/1/014} {\bibfield  {journal} {\bibinfo
  {journal} {Smart Mater. Struct.},\ }\textbf {\bibinfo {volume} {15}},\
  \bibinfo {pages} {S} (\bibinfo {year} {2006})}\BibitemShut {NoStop}%
\bibitem [{\citenamefont {Evans}\ \emph {et~al.}(2008)\citenamefont {Evans},
  \citenamefont {McGlynn}, \citenamefont {Towlson}, \citenamefont {Gunn},
  \citenamefont {Jones}, \citenamefont {Jenkins}, \citenamefont {Winter},\ and\
  \citenamefont {Poolton}}]{Evans2008}%
  \BibitemOpen
  \bibfield  {author} {\bibinfo {author} {\bibfnamefont {D.~A.}\ \bibnamefont
  {Evans}}, \bibinfo {author} {\bibfnamefont {A.~G.}\ \bibnamefont {McGlynn}},
  \bibinfo {author} {\bibfnamefont {B.~M.}\ \bibnamefont {Towlson}}, \bibinfo
  {author} {\bibfnamefont {M.}~\bibnamefont {Gunn}}, \bibinfo {author}
  {\bibfnamefont {D.}~\bibnamefont {Jones}}, \bibinfo {author} {\bibfnamefont
  {T.~E.}\ \bibnamefont {Jenkins}}, \bibinfo {author} {\bibfnamefont
  {R.}~\bibnamefont {Winter}}, \ and\ \bibinfo {author} {\bibfnamefont
  {N.~R.~J.}\ \bibnamefont {Poolton}},\ }\href@noop {} {\bibfield  {journal}
  {\bibinfo  {journal} {Journal of Physics: Condensed Matter},\ }\textbf
  {\bibinfo {volume} {20}},\ \bibinfo {pages} {075233} (\bibinfo {year}
  {2008})}\BibitemShut {NoStop}%
\bibitem [{\citenamefont {Berger}(1997)}]{Berger1997}%
  \BibitemOpen
  \bibfield  {author} {\bibinfo {author} {\bibfnamefont {L.~I.}\ \bibnamefont
  {Berger}},\ }\href@noop {} {\emph {\bibinfo {title} {Semiconductor
  Materials}}}\ (\bibinfo  {publisher} {CRC Press},\ \bibinfo {address} {Boca
  Raton, FL},\ \bibinfo {year} {1997})\BibitemShut {NoStop}%
\bibitem [{\citenamefont {Zakharov}\ \emph {et~al.}(1994)\citenamefont
  {Zakharov}, \citenamefont {Rubio}, \citenamefont {Blase}, \citenamefont
  {Cohen},\ and\ \citenamefont {Louie}}]{Zakharov1994}%
  \BibitemOpen
  \bibfield  {author} {\bibinfo {author} {\bibfnamefont {O.}~\bibnamefont
  {Zakharov}}, \bibinfo {author} {\bibfnamefont {A.}~\bibnamefont {Rubio}},
  \bibinfo {author} {\bibfnamefont {X.}~\bibnamefont {Blase}}, \bibinfo
  {author} {\bibfnamefont {M.~L.}\ \bibnamefont {Cohen}}, \ and\ \bibinfo
  {author} {\bibfnamefont {S.~G.}\ \bibnamefont {Louie}},\ }\Doi
  {10.1103/PhysRevB.50.10780} {\bibfield  {journal} {\bibinfo  {journal} {Phys.
  Rev. B},\ }\textbf {\bibinfo {volume} {50}},\ \bibinfo {pages} {10780}
  (\bibinfo {year} {1994})}\BibitemShut {NoStop}%
\bibitem [{PBE()}]{PBEsol2}%
  \BibitemOpen
  \href@noop {} {}\bibinfo {note} {While PBEsol performed well for the 18
  solids studied, LSDA remained ``unsurpassed'' for semiconductors. See
  Reference~\cite{PBEsol}, page 136406-4, first paragraph.}\BibitemShut {Stop}%
\bibitem [{\citenamefont {Sun}\ \emph {et~al.}(2011)\citenamefont {Sun},
  \citenamefont {Marsman}, \citenamefont {Csonka}, \citenamefont {Ruzsinszky},
  \citenamefont {Hao}, \citenamefont {Kim}, \citenamefont {Kresse},\ and\
  \citenamefont {Perdew}}]{Sun2011}%
  \BibitemOpen
  \bibfield  {author} {\bibinfo {author} {\bibfnamefont {J.}~\bibnamefont
  {Sun}}, \bibinfo {author} {\bibfnamefont {M.}~\bibnamefont {Marsman}},
  \bibinfo {author} {\bibfnamefont {G.~I.}\ \bibnamefont {Csonka}}, \bibinfo
  {author} {\bibfnamefont {A.}~\bibnamefont {Ruzsinszky}}, \bibinfo {author}
  {\bibfnamefont {P.}~\bibnamefont {Hao}}, \bibinfo {author} {\bibfnamefont
  {Y.-S.}\ \bibnamefont {Kim}}, \bibinfo {author} {\bibfnamefont
  {G.}~\bibnamefont {Kresse}}, \ and\ \bibinfo {author} {\bibfnamefont {J.~P.}\
  \bibnamefont {Perdew}},\ }\Doi {10.1103/PhysRevB.84.035117} {\bibfield
  {journal} {\bibinfo  {journal} {Phys. Rev. B},\ }\textbf {\bibinfo {volume}
  {84}},\ \bibinfo {pages} {035117} (\bibinfo {year} {2011})}\BibitemShut
  {NoStop}%
\bibitem [{\citenamefont {Armiento}\ and\ \citenamefont
  {Mattsson}(2005)}]{AM05}%
  \BibitemOpen
  \bibfield  {author} {\bibinfo {author} {\bibfnamefont {R.}~\bibnamefont
  {Armiento}}\ and\ \bibinfo {author} {\bibfnamefont {A.~E.}\ \bibnamefont
  {Mattsson}},\ }\Doi {10.1103/PhysRevB.72.085108} {\bibfield  {journal}
  {\bibinfo  {journal} {Phys. Rev. B},\ }\textbf {\bibinfo {volume} {72}},\
  \bibinfo {pages} {085108} (\bibinfo {year} {2005})}\BibitemShut {NoStop}%
\bibitem [{\citenamefont {Perdew}\ \emph
  {et~al.}(2009){\natexlab{a}}\citenamefont {Perdew}, \citenamefont
  {Ruzsinszky}, \citenamefont {Csonka}, \citenamefont {Constantin},\ and\
  \citenamefont {Sun}}]{revTPSS}%
  \BibitemOpen
  \bibfield  {author} {\bibinfo {author} {\bibfnamefont {J.~P.}\ \bibnamefont
  {Perdew}}, \bibinfo {author} {\bibfnamefont {A.}~\bibnamefont {Ruzsinszky}},
  \bibinfo {author} {\bibfnamefont {G.~I.}\ \bibnamefont {Csonka}}, \bibinfo
  {author} {\bibfnamefont {L.~A.}\ \bibnamefont {Constantin}}, \ and\ \bibinfo
  {author} {\bibfnamefont {J.}~\bibnamefont {Sun}},\ }\Doi
  {10.1103/PhysRevLett.103.026403} {\bibfield  {journal} {\bibinfo  {journal}
  {Phys. Rev. Lett.},\ }\textbf {\bibinfo {volume} {103}},\ \bibinfo {pages}
  {026403} (\bibinfo {year} {2009}{\natexlab{a}})},\ \bibinfo {note} {see also
  the Erratum\cite{revTPSSerr}, Phys. Rev. Lett., \textbf{106}, 179902
  (2011).}\BibitemShut {Stop}%
\bibitem [{\citenamefont {Mori-S\'anchez}\ \emph {et~al.}(2009)\citenamefont
  {Mori-S\'anchez}, \citenamefont {Cohen},\ and\ \citenamefont
  {Yang}}]{MCY2009}%
  \BibitemOpen
  \bibfield  {author} {\bibinfo {author} {\bibfnamefont {P.}~\bibnamefont
  {Mori-S\'anchez}}, \bibinfo {author} {\bibfnamefont {A.~J.}\ \bibnamefont
  {Cohen}}, \ and\ \bibinfo {author} {\bibfnamefont {W.}~\bibnamefont {Yang}},\
  }\href@noop {} {\bibfield  {journal} {\bibinfo  {journal} {Phys. Rev.
  Lett.},\ }\textbf {\bibinfo {volume} {102}},\ \bibinfo {pages} {066403}
  (\bibinfo {year} {2009})}\BibitemShut {NoStop}%
\bibitem [{\citenamefont {Brothers}\ \emph {et~al.}(2008)\citenamefont
  {Brothers}, \citenamefont {Izmaylov}, \citenamefont {Normand}, \citenamefont
  {Barone},\ and\ \citenamefont {Scuseria}}]{Brothers2008}%
  \BibitemOpen
  \bibfield  {author} {\bibinfo {author} {\bibfnamefont {E.~N.}\ \bibnamefont
  {Brothers}}, \bibinfo {author} {\bibfnamefont {A.~F.}\ \bibnamefont
  {Izmaylov}}, \bibinfo {author} {\bibfnamefont {J.~O.}\ \bibnamefont
  {Normand}}, \bibinfo {author} {\bibfnamefont {V.}~\bibnamefont {Barone}}, \
  and\ \bibinfo {author} {\bibfnamefont {G.~E.}\ \bibnamefont {Scuseria}},\
  }\href@noop {} {\bibfield  {journal} {\bibinfo  {journal} {J. Chem. Phys.},\
  }\textbf {\bibinfo {volume} {129}},\ \bibinfo {pages} {011102} (\bibinfo
  {year} {2008})}\BibitemShut {NoStop}%
\bibitem [{\citenamefont {Prodan}\ \emph {et~al.}(2007)\citenamefont {Prodan},
  \citenamefont {Scuseria},\ and\ \citenamefont {Martin}}]{Prodan2007}%
  \BibitemOpen
  \bibfield  {author} {\bibinfo {author} {\bibfnamefont {I.~D.}\ \bibnamefont
  {Prodan}}, \bibinfo {author} {\bibfnamefont {G.~E.}\ \bibnamefont
  {Scuseria}}, \ and\ \bibinfo {author} {\bibfnamefont {R.~L.}\ \bibnamefont
  {Martin}},\ }\href@noop {} {\bibfield  {journal} {\bibinfo  {journal} {Phys.
  Rev. B},\ }\textbf {\bibinfo {volume} {76}},\ \bibinfo {pages} {033103}
  (\bibinfo {year} {2007})}\BibitemShut {NoStop}%
\bibitem [{\citenamefont {Prodan}\ \emph {et~al.}(2006)\citenamefont {Prodan},
  \citenamefont {Scuseria},\ and\ \citenamefont {Martin}}]{Prodan2006}%
  \BibitemOpen
  \bibfield  {author} {\bibinfo {author} {\bibfnamefont {I.~D.}\ \bibnamefont
  {Prodan}}, \bibinfo {author} {\bibfnamefont {G.~E.}\ \bibnamefont
  {Scuseria}}, \ and\ \bibinfo {author} {\bibfnamefont {R.~L.}\ \bibnamefont
  {Martin}},\ }\href@noop {} {\bibfield  {journal} {\bibinfo  {journal} {Phys.
  Rev. B},\ }\textbf {\bibinfo {volume} {73}},\ \bibinfo {pages} {045104}
  (\bibinfo {year} {2006})}\BibitemShut {NoStop}%
\bibitem [{\citenamefont {Prodan}\ \emph {et~al.}(2005)\citenamefont {Prodan},
  \citenamefont {Sordo}, \citenamefont {Kudin}, \citenamefont {Scuseria},\ and\
  \citenamefont {Martin}}]{Prodan2005}%
  \BibitemOpen
  \bibfield  {author} {\bibinfo {author} {\bibfnamefont {I.~D.}\ \bibnamefont
  {Prodan}}, \bibinfo {author} {\bibfnamefont {J.~A.}\ \bibnamefont {Sordo}},
  \bibinfo {author} {\bibfnamefont {K.~N.}\ \bibnamefont {Kudin}}, \bibinfo
  {author} {\bibfnamefont {G.~E.}\ \bibnamefont {Scuseria}}, \ and\ \bibinfo
  {author} {\bibfnamefont {R.~L.}\ \bibnamefont {Martin}},\ }\href@noop {}
  {\bibfield  {journal} {\bibinfo  {journal} {J. Chem. Phys.},\ }\textbf
  {\bibinfo {volume} {123}},\ \bibinfo {pages} {014703} (\bibinfo {year}
  {2005})}\BibitemShut {NoStop}%
\bibitem [{\citenamefont {Alibert}\ \emph {et~al.}(1983)\citenamefont
  {Alibert}, \citenamefont {Joullie}, \citenamefont {Joullie},\ and\
  \citenamefont {Ance}}]{Alibert1983}%
  \BibitemOpen
  \bibfield  {author} {\bibinfo {author} {\bibfnamefont {C.}~\bibnamefont
  {Alibert}}, \bibinfo {author} {\bibfnamefont {A.}~\bibnamefont {Joullie}},
  \bibinfo {author} {\bibfnamefont {A.~M.}\ \bibnamefont {Joullie}}, \ and\
  \bibinfo {author} {\bibfnamefont {C.}~\bibnamefont {Ance}},\ }\href@noop {}
  {\bibfield  {journal} {\bibinfo  {journal} {Phys. Rev. B},\ }\textbf
  {\bibinfo {volume} {27}},\ \bibinfo {pages} {4946} (\bibinfo {year}
  {1983})}\BibitemShut {NoStop}%
\bibitem [{\citenamefont {Matthieu}\ \emph {et~al.}(1975)\citenamefont
  {Matthieu}, \citenamefont {Auvergne}, \citenamefont {Merle},\ and\
  \citenamefont {Rustagi}}]{Mathieu1975}%
  \BibitemOpen
  \bibfield  {author} {\bibinfo {author} {\bibfnamefont {H.}~\bibnamefont
  {Matthieu}}, \bibinfo {author} {\bibfnamefont {D.}~\bibnamefont {Auvergne}},
  \bibinfo {author} {\bibfnamefont {P.}~\bibnamefont {Merle}}, \ and\ \bibinfo
  {author} {\bibfnamefont {K.~C.}\ \bibnamefont {Rustagi}},\ }\href@noop {}
  {\bibfield  {journal} {\bibinfo  {journal} {Phys. Rev. B},\ }\textbf
  {\bibinfo {volume} {12}},\ \bibinfo {pages} {5846} (\bibinfo {year}
  {1975})}\BibitemShut {NoStop}%
\bibitem [{\citenamefont {Joullie}\ \emph {et~al.}(1982)\citenamefont
  {Joullie}, \citenamefont {Girault}, \citenamefont {Joullie},\ and\
  \citenamefont {Zien-Eddine}}]{Joullie1982}%
  \BibitemOpen
  \bibfield  {author} {\bibinfo {author} {\bibfnamefont {A.}~\bibnamefont
  {Joullie}}, \bibinfo {author} {\bibfnamefont {B.}~\bibnamefont {Girault}},
  \bibinfo {author} {\bibfnamefont {A.~M.}\ \bibnamefont {Joullie}}, \ and\
  \bibinfo {author} {\bibfnamefont {A.}~\bibnamefont {Zien-Eddine}},\
  }\href@noop {} {\bibfield  {journal} {\bibinfo  {journal} {Phys. Rev. B},\
  }\textbf {\bibinfo {volume} {25}},\ \bibinfo {pages} {7830} (\bibinfo {year}
  {1982})}\BibitemShut {NoStop}%
\bibitem [{\citenamefont {Grundmann}(2010){\natexlab{d}}}]{Grundmann2010}%
  \BibitemOpen
  \bibfield  {author} {\bibinfo {author} {\bibfnamefont {M.}~\bibnamefont
  {Grundmann}},\ }\href@noop {} {\emph {\bibinfo {title} {Physics of
  Semiconductors}}}\ (\bibinfo  {publisher} {Springer},\ \bibinfo {address}
  {Berlin},\ \bibinfo {year} {2010})\BibitemShut {NoStop}%
\bibitem [{\citenamefont {Safa~Kasap}(2006)}]{Springer2006}%
  \BibitemOpen
  \bibfield  {author} {\bibinfo {author} {\bibfnamefont {P.~C.~E.}\
  \bibnamefont {Safa~Kasap}},\ }\href@noop {} {\emph {\bibinfo {title}
  {Springer Handbook of Electronic and Photonic Materials}}}\ (\bibinfo
  {publisher} {Springer},\ \bibinfo {address} {Berlin},\ \bibinfo {year}
  {2006})\BibitemShut {NoStop}%
\bibitem [{\citenamefont {Economou}(2010)}]{Economou2010}%
  \BibitemOpen
  \bibfield  {author} {\bibinfo {author} {\bibfnamefont {E.~N.}\ \bibnamefont
  {Economou}},\ }\href@noop {} {\emph {\bibinfo {title} {The Physics of
  Semiconductors Essentials and Beyond}}}\ (\bibinfo  {publisher} {Springer},\
  \bibinfo {address} {Berlin},\ \bibinfo {year} {2010})\BibitemShut {NoStop}%
\bibitem [{\citenamefont {Song}\ \emph {et~al.}(2011)\citenamefont {Song},
  \citenamefont {Yamashita},\ and\ \citenamefont {Hirao}}]{Hirao2011}%
  \BibitemOpen
  \bibfield  {author} {\bibinfo {author} {\bibfnamefont {J.-W.}\ \bibnamefont
  {Song}}, \bibinfo {author} {\bibfnamefont {K.}~\bibnamefont {Yamashita}}, \
  and\ \bibinfo {author} {\bibfnamefont {K.}~\bibnamefont {Hirao}},\
  }\href@noop {} {\bibfield  {journal} {\bibinfo  {journal} {J. Chem. Phys.},\
  }\textbf {\bibinfo {volume} {135}},\ \bibinfo {pages} {071103} (\bibinfo
  {year} {2011})}\BibitemShut {NoStop}%
\bibitem [{\citenamefont {Marsman}\ \emph {et~al.}(2008)\citenamefont
  {Marsman}, \citenamefont {Paier}, \citenamefont {Stroppa},\ and\
  \citenamefont {Kresse}}]{MPSK2008}%
  \BibitemOpen
  \bibfield  {author} {\bibinfo {author} {\bibfnamefont {M.}~\bibnamefont
  {Marsman}}, \bibinfo {author} {\bibfnamefont {J.}~\bibnamefont {Paier}},
  \bibinfo {author} {\bibfnamefont {A.}~\bibnamefont {Stroppa}}, \ and\
  \bibinfo {author} {\bibfnamefont {G.}~\bibnamefont {Kresse}},\ }\href@noop {}
  {\bibfield  {journal} {\bibinfo  {journal} {Journal of Physics: Condensed
  Matter},\ }\textbf {\bibinfo {volume} {20}},\ \bibinfo {pages} {064201}
  (\bibinfo {year} {2008})}\BibitemShut {NoStop}%
\bibitem [{\citenamefont {El-Mellouhi}\ \emph {et~al.}(2011)\citenamefont
  {El-Mellouhi}, \citenamefont {Brothers}, \citenamefont {Lucero},\ and\
  \citenamefont {Scuseria}}]{Fadwa2011}%
  \BibitemOpen
  \bibfield  {author} {\bibinfo {author} {\bibfnamefont {F.}~\bibnamefont
  {El-Mellouhi}}, \bibinfo {author} {\bibfnamefont {E.~N.}\ \bibnamefont
  {Brothers}}, \bibinfo {author} {\bibfnamefont {M.~J.}\ \bibnamefont
  {Lucero}}, \ and\ \bibinfo {author} {\bibfnamefont {G.~E.}\ \bibnamefont
  {Scuseria}},\ }\href@noop {} {\bibfield  {journal} {\bibinfo  {journal}
  {Phys. Rev. B.},\ }\textbf {\bibinfo {volume} {83}},\ \bibinfo {pages}
  {205128} (\bibinfo {year} {2011})}\BibitemShut {NoStop}%
\bibitem [{\citenamefont {Snure}\ and\ \citenamefont
  {Tiwari}(2007)}]{SnureCuBO2}%
  \BibitemOpen
  \bibfield  {author} {\bibinfo {author} {\bibfnamefont {M.}~\bibnamefont
  {Snure}}\ and\ \bibinfo {author} {\bibfnamefont {A.}~\bibnamefont {Tiwari}},\
  }\href@noop {} {\bibfield  {journal} {\bibinfo  {journal} {Appl. Phys.
  Lett.},\ }\textbf {\bibinfo {volume} {91}},\ \bibinfo {pages} {092123}
  (\bibinfo {year} {2007})}\BibitemShut {NoStop}%
\bibitem [{\citenamefont {Scanlon}\ \emph {et~al.}(2009)\citenamefont
  {Scanlon}, \citenamefont {Walsh},\ and\ \citenamefont
  {Watson}}]{ScanlonCuBO2}%
  \BibitemOpen
  \bibfield  {author} {\bibinfo {author} {\bibfnamefont {D.~O.}\ \bibnamefont
  {Scanlon}}, \bibinfo {author} {\bibfnamefont {A.}~\bibnamefont {Walsh}}, \
  and\ \bibinfo {author} {\bibfnamefont {G.~W.}\ \bibnamefont {Watson}},\
  }\href@noop {} {\bibfield  {journal} {\bibinfo  {journal} {Chem. Mater.},\
  }\textbf {\bibinfo {volume} {21}},\ \bibinfo {pages} {4568} (\bibinfo {year}
  {2009})}\BibitemShut {NoStop}%
\bibitem [{\citenamefont {Trani}\ \emph {et~al.}(2010)\citenamefont {Trani},
  \citenamefont {Vidal}, \citenamefont {Botti},\ and\ \citenamefont
  {Marques}}]{TraniCuBO2}%
  \BibitemOpen
  \bibfield  {author} {\bibinfo {author} {\bibfnamefont {F.}~\bibnamefont
  {Trani}}, \bibinfo {author} {\bibfnamefont {J.}~\bibnamefont {Vidal}},
  \bibinfo {author} {\bibfnamefont {S.}~\bibnamefont {Botti}}, \ and\ \bibinfo
  {author} {\bibfnamefont {M.~A.~L.}\ \bibnamefont {Marques}},\ }\Doi
  {10.1103/PhysRevB.82.085115} {\bibfield  {journal} {\bibinfo  {journal}
  {Phys. Rev. B},\ }\textbf {\bibinfo {volume} {82}},\ \bibinfo {pages}
  {085115} (\bibinfo {year} {2010})}\BibitemShut {NoStop}%
\bibitem [{\citenamefont {Heyd}\ \emph {et~al.}(2006)\citenamefont {Heyd},
  \citenamefont {Scuseria},\ and\ \citenamefont {Ernzerhof}}]{HSE06e}%
  \BibitemOpen
  \bibfield  {author} {\bibinfo {author} {\bibfnamefont {J.}~\bibnamefont
  {Heyd}}, \bibinfo {author} {\bibfnamefont {G.~E.}\ \bibnamefont {Scuseria}},
  \ and\ \bibinfo {author} {\bibfnamefont {M.}~\bibnamefont {Ernzerhof}},\
  }\href@noop {} {\bibfield  {journal} {\bibinfo  {journal} {J. Chem. Phys.},\
  }\textbf {\bibinfo {volume} {124}},\ \bibinfo {pages} {219906} (\bibinfo
  {year} {2006})}\BibitemShut {NoStop}%
\bibitem [{\citenamefont {Perdew}\ \emph
  {et~al.}(2009){\natexlab{b}}\citenamefont {Perdew}, \citenamefont
  {Ruzsinszky}, \citenamefont {Ernzerhof}, \citenamefont {Csonka},
  \citenamefont {Vydrov}, \citenamefont {Scuseria}, \citenamefont {Constantin},
  \citenamefont {Zhou},\ and\ \citenamefont {Burke}}]{PBEsolErr}%
  \BibitemOpen
  \bibfield  {author} {\bibinfo {author} {\bibfnamefont {J.~P.}\ \bibnamefont
  {Perdew}}, \bibinfo {author} {\bibfnamefont {A.}~\bibnamefont {Ruzsinszky}},
  \bibinfo {author} {\bibfnamefont {M.}~\bibnamefont {Ernzerhof}}, \bibinfo
  {author} {\bibfnamefont {G.~I.}\ \bibnamefont {Csonka}}, \bibinfo {author}
  {\bibfnamefont {O.~A.}\ \bibnamefont {Vydrov}}, \bibinfo {author}
  {\bibfnamefont {G.~E.}\ \bibnamefont {Scuseria}}, \bibinfo {author}
  {\bibfnamefont {L.~A.}\ \bibnamefont {Constantin}}, \bibinfo {author}
  {\bibfnamefont {X.}~\bibnamefont {Zhou}}, \ and\ \bibinfo {author}
  {\bibfnamefont {K.}~\bibnamefont {Burke}},\ }\Doi
  {10.1103/PhysRevLett.102.039902} {\bibfield  {journal} {\bibinfo  {journal}
  {Phys. Rev. Lett.},\ }\textbf {\bibinfo {volume} {102}},\ \bibinfo {pages}
  {039902} (\bibinfo {year} {2009}{\natexlab{b}})}\BibitemShut {NoStop}%
\bibitem [{\citenamefont {Perdew}\ \emph {et~al.}(2011)\citenamefont {Perdew},
  \citenamefont {Ruzsinszky}, \citenamefont {Csonka}, \citenamefont
  {Constantin},\ and\ \citenamefont {Sun}}]{revTPSSerr}%
  \BibitemOpen
  \bibfield  {author} {\bibinfo {author} {\bibfnamefont {J.~P.}\ \bibnamefont
  {Perdew}}, \bibinfo {author} {\bibfnamefont {A.}~\bibnamefont {Ruzsinszky}},
  \bibinfo {author} {\bibfnamefont {G.~I.}\ \bibnamefont {Csonka}}, \bibinfo
  {author} {\bibfnamefont {L.~A.}\ \bibnamefont {Constantin}}, \ and\ \bibinfo
  {author} {\bibfnamefont {J.}~\bibnamefont {Sun}},\ }\href@noop {} {\bibfield
  {journal} {\bibinfo  {journal} {Phys. Rev. Lett.},\ }\textbf {\bibinfo
  {volume} {106}},\ \bibinfo {pages} {179902} (\bibinfo {year}
  {2011})}\BibitemShut {NoStop}%
\end{thebibliography}%

\end{document}